\documentclass[twocolumn,floatfix,superscriptaddress]{revtex4}
\usepackage{graphicx}
\usepackage{amsmath}
\usepackage{latexsym}
\usepackage{epstopdf}
\usepackage{amsmath}
\usepackage{amssymb,amsmath}
\usepackage[toc,page]{appendix}
\newcommand{\beq}{\begin{equation}}
\newcommand{\eeq}{\end{equation}}

\begin{document}

\title{Assessment of 48 Stock markets using adaptive multifractal approach}

\author{Paulo Ferreira}
\affiliation{CEFAGE-UE, IIFA, Universidade de \'{E}vora, Largo dos Colegiais 2, 7000 \'{E}vora, Portugal
Departamento de Ci\^{e}ncias Agr\'{a}rias e Veterin\'{a}rias, Escola Superior Agr\'{a}ria de Elvas, Instituto Polit\^{e}cnico de Portalegre, Portugal}
\affiliation{Universidade Europeia, Laureate International Universities, 1500-210 Lisboa, Portugal}
\author{Andreia Dion\'{\i}sio}
\affiliation{CEFAGE-UE (Center for Advanced Studies in Management and Economics of the University of \'{E}vora)}

\author{ S.M.S. Movahed }
\affiliation{Department of Physics, Shahid Beheshti University, Velenjak, Tehran 19839, Iran}
\affiliation{School of Physics, Institute for researches in fundamental sciences, P.O.Box 19395-5531,Tehran,
Iran}

\vskip 1cm

\begin{abstract}

In this paper,  Stock market comovements are examined using cointegration, Granger causality tests and nonlinear approaches in context of  mutual information and correlations. Since underlying data sets are affected by non-stationarities and trends, we also apply Adaptive Multifractal
Detrended Fluctuation Analysis (AMF-DFA) and Adaptive Multifractal Detrended
Cross-Correlation Analysis (AMF-DXA). We find only 170 pair
of Stock markets cointegrated, and according to the Granger
causality and mutual information, we realize that the strongest
relations lies between emerging markets, and between emerging and
frontier markets. According to scaling exponent given by AMF-DFA,
$h(q=2)>1$, we find that all underlying data sets belong to
non-stationary process. According to Efficient Market Hypothesis (EMH), only  8 markets are classified in uncorrelated processes at $2\sigma$ confidence interval. 6 Stock markets belong to anti-correlated class and dominant part of markets has memory in corresponding daily index prices during January 1995 to February 2014.  New-Zealand with $H=0.457\pm0.004$ and Jordan with $H=0.602\pm 0.006$ are far from EMH. The nature of cross-correlation exponents based on AMF-DXA is almost
multifractal for all pair of Stock markets. The empirical relation,
$H_{xy}\le [H_{xx}+H_{yy}]/2$, is confirmed. Mentioned relation for $q>0$ is also satisfied 
while for $q<0$ there is a deviation from this relation confirming behavior of markets for small fluctuations is affected by contribution of major pair. For larger fluctuations, the cross-correlation contains information from both local (internal) and global (external) conditions. Width of singularity spectrum for auto-correlation and cross-correlation are $\Delta \alpha_{xx}\in [0.304,0.905]$ and $\Delta \alpha_{xy}\in [0.246,1.178]$, respectively. The wide range of singularity spectrum for cross-correlation confirms that the bilateral relation between Stock markets is more complex. The value
of $\sigma_{DCCA}$ indicates that all pairs of stock market  studied in this time
interval belong to cross-correlated processes.\\
Keyword: long-range relationship,  Stock markets, cointegration,
mutual information, detrended cross-correlation analysis, multifractal, adaptive detrending.
\end{abstract}
\maketitle

\section{Introduction}

he assessment of time series is one topic of major interest in
economics, management and econophysics. Risk and optimal portfolio managements as well as quantifying volatility in various markets need to examine different aspect of markets \cite{peter03}. Since, such kind of systems have complex parts, subsequently statistical approaches enable us to mitigate risk in trading. One of important tasks to do in an analysis of financial markets, is examining, whether temporal or sectional, could lead to any prediction of the
series and the possibility of violating the assumption of efficient
markets. Therefore, the notion of Efficient Markets Hypothesis (EMH) is a 
critical subject in this field. In fact, a financial market is considered
efficient in its weak form if it is not possible to identify any
deterministic pattern in its time series behavior. This means that
there is no possibility, through arbitrage, of obtaining systematic
abnormal profits using past information \cite{fama70}. In the light
of this theory, financial markets have been subjected to extensive
analysis to check whether there are windows of profit opportunities,
considering the fluctuations and dynamics of markets themselves \cite{pagan90}. One of the first studies applied to the behavior of
financial markets series studied the probability distribution of share prices and
concluded that prices follow a Gaussian distribution \cite{bach900}, other studies confirmed that Stock prices are randomly determined \cite{kenda53}.
Some other studies, validated the random walk hypothesis, which seems to indicate that asset prices
have no memory and are therefore independent in time \cite{osbo64,morge64,fama63}. For a long
time Bachelier's theory  \cite{bach900} that financial series behave
like a random walk was accepted and introduced in many economic
models, such as the efficient markets hypothesis \cite{fama70}. However, several studies contradicted this evidence,
finding the existence of stylized facts \cite{cont01}. One of these stylized facts is the existence of
fat tails in returns distributions, which is related to the fact
that the volatility of assets returns is higher than expected by a
Gaussian distribution \cite{osbo64}. Other stylized
facts that may contribute to reject the evidence of normality in
assets returns are found, including the existence of asymmetries in gains and
losses (loss movements are more pronounced); greater than the
expected intermittency and variability of returns, with volatility
clustering behavior; leverage effect (negative relation between
volatility and profitability); correlation between trading volumes
and volatility; and existence of autocorrelation in variance \cite{cont01}. 

The
analysis of serial dependence, both linear and nonlinear, has some
relevances in the financial literature in recent times. In most
cases, empirical studies identified the possibility of
autocorrelation. However, generally these linear autocorrelations
quickly disappear, although there are authors defending the
existence of long-term dependence \cite{camp87}. Motivated by appearing emergent behavior in some physical systems, mutual influence in the context of bilateral and multilateral relations and stochastic effect of other markets on underlying system, indeed a vital purpose in the frame work of beyond one-point statistics. Therefore, beside trivial assessments, finding cross-correlation using different approaches not only provides a reliable strategy to determine degree of cross-correlation, but also make an opportunity to recognize temporal and spatial correlation accompanying necessity information to establish optimal portfolio management. In order to produce mentioned objectives in econometrics, there are many researches have been done.  According to so-called level crossing analysis, a new criterion was introduced to quantify degree of development in the Stock markets \cite{jafarilevel}.  Recently, in order to determine states in Stock markets, Michael C. M\"{u}nnix et. al., proposed a new method to identify the states of financial markets \cite{munix12}. They examined the temporal cross-correlation of financial data from S\&P 500 Stocks and could categorize  the states of underlying data sets. Such analysis can be used for risk management \cite{munix12}.

Relevant studies on market efficiency used linear equations to
analyze return rate dependence, failing to detect other types of
dependency, including nonlinear dependence. Therefore, in order to
study financial markets, it is necessary to follow general models
which are capable of capturing global, and not only linear,
dependence. In this context, mutual information was introduced and
its properties were explored as a measure of dependence in time
series. This method has some advantages, because it considers the
whole structure of time series, linear and nonlinear \cite{darb00}
and potentially can be considered for estimation of financial risk
\cite{munix12}. Methods to analyze long-term
dependence in time series have been developed, like detrended
fluctuation analysis (DFA) and its generalized version  \cite{peng92,peng94,bul95,bun02} and
detrended cross-correlation analysis (DCCA) \cite{DCCA}. DFA studies
the behavior of individual series while DCCA is a methodology which
analyzes the behavior of pairs of series. If data are affected by global and local trends and noise, one should apply robust detrending methods to clean data for further applications. In principle trends are classified in two categories: Global and local trends. Some trivial approaches to remove global trends are considering an arbitrary polynomial function or/and empirical mode decomposition \cite {wu07}. Local trends based on smoothing and segmentation  techniques are popular but it has been demonstrated that they are not very accurate \cite{hu09}. 
When cross-correlation analysis is supposed to investigate, utilizing proper detrending algorithm has important impact on associated results.  To this end, local and global detrending approaches in cross-correlation analysis have been investigated in \cite{DCCA2}. Power-law cross-correlated processes was considered in \cite{DCCA3}. Influence of external forces has also elucidated in \cite{DCCA4}.  A new approach introduced by G.F. Zebende \cite{zeb11} to quantify the cross-correlation coefficient based on DCCA is another useful method (see also \cite{Zebende13}). 

\begin{table}
\begin{center}
\scalebox{0.7}{
\begin{tabular}{|c|c|c|c|c|c|}
  \hline
      Index&Country name & Index&Country name  & Index&Country name   \\\hline
   1&ARGENTINA & 17&HONG-KONG &33& PAKISTAN \\\hline
  2& AUSTRALIA &  18& HUNGARY    &34& PERU                                                \\\hline
  3&  AUSTRIA &19&INDIA &35&POLAND \\\hline
    4& BELGIUM &20& INDONESIA & 36& PORTUGAL\\\hline
    5&  BRAZIL & 21&IRELAND &37&RUSSIA\\\hline
 6&    CANADA &22& ISRAEL-L  &38&SINGAPORE\\\hline
 7& CHILE & 23& ITALY&39&SOUTH-AFRICA\\\hline
 8&CHINA &24& JAPAN & 40 &SPAIN\\\hline
9&COLOMBIA & 25& JORDAN& 41& SRI-LANKA\\\hline
10&CZECH-REPUBLIC &26&  KOREA& 42& SWEDEN\\\hline
11&DENMARK & 27& MALAYSIA&43& SWITZERLAND\\\hline
12&EGYPT & 28& MEXICO& 44& TAIWAN\\\hline
13&FINLAND &29& MOROCCO& 45& THAILAND \\\hline
14&FRANCE& 30& NETHERLANDS& 46&TURKEY\\\hline
15&GREECE & 31&NEW-ZEALAND& 47& UK\\\hline
16&GERMANY&32&NORWAY& 48&USA \\\hline
\end{tabular}}
\caption{\label{index1}The name of countries and their indices used in this paper.}
\end{center}
\end{table}

DCCA, like DFA, is a
methodology original from Physics but it is also applied to
economics and, in particular, to financial markets. It was used, for example, to
analyze the behavior between price and volume change in several
indices which found a cross-correlation between these variables,
indicating that price changes depend on previous changes but also on
volume changes \cite{pod09}. Recently Shi et. al., introduced a multiscale multifractal detrended cross-correlation analysis and applied it for analyzing some financial indices \cite{shi14}. Another work analyzed the behavior of
the Stock returns using daily data of six indices: three American
and three Chinese indices, from 2002 to 2009, and also found a
power-law cross-correlation between data \cite{lin09}. We can find different works in different areas such as the analysis of data of
agricultural future markets between China and USA \cite{chen11}, the
analysis of the behavior between Chinese and surrounding Stock
markets \cite{ma13} or the  use information about financial markets
in China \cite{cao14}. The three works previously indicated show
that emerging markets have some attention from researchers. Another more recent researches on Stock markets based on multiscale analysis and multifractal cross-correlation methods have been done by A. Lin et. al., \cite{lin14} and X. Zhao et. al., \cite{zhao14}. In the former paper, interactions and structure of US and China Stock markets have been examined while in the latter work, the authors  investigated the spectrum of multifractality in cross-correlation analysis of china Stock markets. Particularly, DCCA methods have been used in order to examine Oli and US dollar exchange \cite{Reboredo14}. Also, DCCA approach for quantifying cross-correlation between Ibovespa and Brazilian blue-chips has been done in  \cite{Silv15}.

In the present paper, we use several different methodologies to analyse how stock markets behave, regarding to their cross-correlations. We start to use linear methodologies, like cointegration, which allow us to identify possible long run relationships between stock markets. Then, because linear approaches could be not enough, we try to combine cointegration with non-linear approaches. Mutual information is one of those approaches. But the innovative pattern of this paper, is the fact that we try to combine those analyses with an adaptive detrending algorithm to DCCA and its generalized form to apply on 48 Stock markets series which list can be consulted in Table I with following advantages and novelties. Besides this novelty, our paper goes further once it uses a large number of indices, comparing their behaviors. We make the analysis of series, based on unit root, cointegration and Granger causality tests. Then, we complement the analysis with nonlinear approaches in the presence of non-stationarity and trends. In addition to revisiting multi-scale methods, we utilize adaptive detrending algorithm to ensure about the reliability of corresponding computed exponents. Local and global trends embedded in data can be well organized with combination of Adaptive algorithm with DCCA and MF-DXA method \cite{mf-dxa}. Finally, the mutual information of Stock markets will be quantified for our data sets.

The remaining of this paper is organized as follows: Section 2 presents the theoretical background for the methodologies uses: firstly the traditional linear methodologies (unit root, cointegration and Granger causality tests) and next the nonlinear methodologies (mutual information, DFA and DCCA). Data description will be devoted in section 3. Section 4 reports the computational analysis and its results and Section 5 gives discussions and conclusions.

\section{Background theory}
In this section we introduce most relevant techniques to analyze of Stock market data used as input series.

\subsection{Unit root, cointegration and Granger causality analysis}

We are starting to analyze the behavior of Stock markets time series, testing the hypothesis of stationarity. Dickey Fuller (or its augmented version) is the most common test to be used. However, its usage should be made carefully because if exist any structural break, results must be misleading. So we use a test which allows to verify the presence or not of stationarity even in the presence of structural breaks \cite{perr92}. This test could be made using one or two structural breaks \cite{clem98}. Due to lack of computational capacity, we chose just one break. This is an appropriate approach once we just want to verify the existence of structural breaks and not to identity its number. This test, which has two different methods (Innovation Outlier and Additive Outlier) has advantages, once if we can reject null hypothesis with structural break, it also can be rejected if it does not exist \cite{perr97}. After testing for unit roots, it can be used Ordinary Least Squares if series are stationary. However, if series are not stationary, we have to study the existence of cointegration between pairs of series. Traditional tests are not appropriated if structural breaks exist \cite{enge87,joha91}. In this case we applied a test with the same null hypothesis of absence of cointegration but allowing for the existence of structural breaks \cite{greg96}. The non-rejection of the previous test implies that the pair of series which is analyzed does not have evidence of long-term relationship. However, if we reject null hypothesis, exist a long-range relationship between those series. In these cases, we performed causality granger tests \cite{greg69}, based on the Vector Error Correlated Method (VECM) result for the first difference of the series.

\subsection{Mutual information}

Mutual information gives the common information between two or more different distributions. Introduced in the literature by Shannon \cite{shan48}, this concept has been improved and widely used over time. In the context of time series, it is used to analyze dependence over time. Mutual information can be understood as a measure of dependence or correlation. However, we should be careful in its interpretation, as it does not provide indication of causality between variables. Mutual information is given by the following expression:
\begin{equation}\label{eq1}
I(X,Y)=\int dx\int dy P_{\rm{Joint}}(x,y)\log\left(\frac{P_{\rm{Joint}}(x,y)}{P_{X}(x)P_{Y}(y)}\right)
\end{equation}
where $P_{\rm{Joint}}(x,y)$ and $P(x)$ are joint and one-point probability density functions, respectively.  $I(X,Y)$ can take any positive value or may be zero. It will be zero if variables are independent (and therefore have no information in common). This makes mutual information an imperfect measure of dependence, since it does not take only absolute values between 0 and 1 \cite{gran04}. It is therefore necessary to standardize it to make direct comparisons \cite{lin94,darb98,soofi97}. One possible normalization is:
\begin{equation}\label{eq2}
\lambda(X,Y)=\sqrt{1-e^{-2I(X,Y)}}
\end{equation}
The measure of dependency identified by Eq. (\ref{eq2}) could vary
between 0 and 1 and can be interpreted as a correlation that is
based on information theory, taking the 0 value if the variables $X$
and $Y$ are independent (i.e. if the variables do not have
information in common). The maximum value is obtained in the case of
a perfect relationship between two variables, i.e., in a
deterministic context. It is used as an alternative to other tests
because it presents several advantages. Firstly, some of the
previous tests have some limitations. For example, the Pearson
correlation coefficient only captures the existence of linear
correlations, but non-linear correlations may also be present in the
data. Thus, mutual information may be used as a measure of overall
correlation and not just of linear correlation. For this reason it
is irrelevant if the sign of the relationship is positive or
negative. Moreover, measures related to entropy require fewer
assumptions and are more flexible. Mutual information is used to
test global dependence of a time series. The null hypothesis is
defined as $H0: I(X,Y) = 0$, meaning that variables are independent
(or that a given time series has no memory). The alternative
hypothesis is given by   $H1: I(X,Y) > 0$. The decision of rejecting
or not rejecting the null hypothesis is made by comparing the
relevant values with the critical ones \cite{dion06}. This test has
the particularity of not needing assumptions on the linearity,
normality or stationarity of time series. However results are more
robust in the case of stationary time series because there is
insufficient evidence of the robustness of this test when
nonlinearity and nonstationarity simultaneously occur
\cite{ferna01}). We estimate mutual information by the
equiquantization method.

\subsection{Adaptive Multifractal Detrended Cross-Correlation Analysis (AMF-DFA)}

When we are going to compare the behavior of financial time series, one of the problems is the possibility of nonstationarity, which avoids using some econometric techniques. Even if series are cointegrated, the results of Ordinary Least Squares cannot be fully interpreted, namely the hypothesis tests to analyze correlation between series.  To this end, traditional methods are encountered with inaccuracies. Jun et. al. have proposed a method for analyzing cross-correlation properties of a series by decomposing the original signal into its positive and negative fluctuation components  \cite{woo}. Based on decomposition of original signal into its positive and negative fluctuation components, recently Podobnik et. al. introduced the cross-correlation between two non-stationary fluctuations, so-called Detrended Cross-Correlation Analysis (DCCA) \cite{DCCA}. After that by means of MF-DFA \cite{bun02} the generalization form of DCCA
 which is called Multifractal Detrended
Cross-Correlation Analysis (MF-DXA) has been elaborated
\cite{mf-dxa}. Mentioned method originally used to explain the
behavior of  natural phenomena, meanwhile both mentioned techniques
could also be applied to economic time series, e.g. financial data.
The existence of trends and non-stationarities in underlying data
causes the correlation exponents become inaccurate and unreliable.
Trends and non-stationarities embedded in measurements  are
generally classified in two categories: Global  and local trends. Beside this classification, one has also two important class for trends from mathematical point of view as: Polynomial and Sinusoidal
trends. In global detrending, linear, polynomial  or exponential  functions are assumed. In Empirical mode decomposition local extrema are determined and by using them the intrinsic trend functions to be computed \cite{wu07}. For local detrending approached, moving average of data is done. In many cases, MF-DFA and MF-DXA methods are not able to remove superimposed trend in data, especially in the presence of sinusoidal
trends \cite{kunhu,trend2}. As the complementary procedure to
eliminate mentioned trend Fourier Detrended Fluctuations Analysis
(F-DFA) \cite{f-dfa,f-dfa2,movahedplasma}, which is actually a high-pass
filter, Singular Value Decomposition (SVD)
\cite{golub,trend3,trend3-1,dccasadegh} and adaptive detrending method \cite{hu09} are recommended to use. Indeed the
most relevant signature for superposition of sinusoidal trends is
the existence of cross-over in the fluctuation function of MF-DFA or
MF-DXA (see following subsection for more details). In this paper we use adaptive detrending algorithm to remove local trends and then MF-DXA method will be applied on clean data sets.  In the
following, we explain the detail of adaptive MF-DXA (AMF-DXA) as a method for analyzing
Stock data sets.

\begin{figure}
\begin{center}
\includegraphics[width=0.7\linewidth]{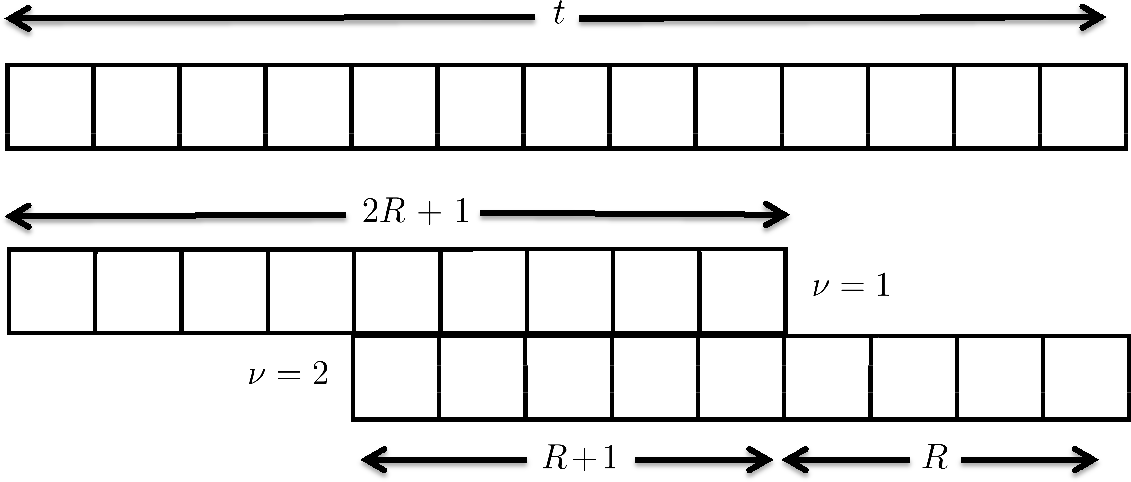}
\caption{\label{adap} Schematic of partitioning a typical series in adaptive detrending method. In each segment, $R+1$ points overlap with neighbour partition. }
\end{center}
\end{figure}
Detrended Cross-Correlation Analysis (DCCA) \cite{DCCA,woo} is a generalization of the DFA \cite{peng92,peng94} method in which only one time
series was analyzed. The adaptive detrending analysis is carried out on the underlying data set to remove local trends. Then, the clean data is used as an input for DCCA and its generalization \cite{hu09}. In order to take into account higher moments,  DCCA
method was modifies and  so-called MF-DXA, has been introduced
\cite{mf-dxa}. The adaptive detrending MF-DXA has
$5$ steps (see \cite{DCCA,hu09,mf-dxa}
for more details):\\
(I): Suppose that an observed equidistance data given by $z_k$ with $k = 1,..., t$. We make segmentations of length $2R+1$. Each neighbouring partitions have $R+1$ overlap points. For each window with size $2R+1$, an arbitrary polynomial ($\mathcal{Y}$) is fitted. Indeed the best polynomial of order $K$ plays corresponding local trend. To construct continuous trend function, following weighted function is defined for overlap part of $\nu$th segment \cite{hu09}:
\begin{eqnarray}
\mathcal{Y}_{\nu}^{{\rm overlap}}(k)=\left(1-\frac{k-1}{R}\right)\mathcal{Y}_{\nu}(k+R)+\frac{k-1}{R}\mathcal{Y}_{\nu+1}(k)
\end{eqnarray}  
where $k=1,2,...,R+1$. The value of $R$ and the order of fitting function are two free parameters should be determined properly \cite{hu09}. In this paper we consider the number of segmentations equal to ${\rm w_{adaptive}}=10$, ${\rm w_{adaptive}}=50$ and ${\rm w_{adaptive}}=100$. Also the second order of fitting polynomial is chosen. The number of points in each segment is calculated by: $2R+1\equiv2\times{\rm int}\left[\frac{t-1}{{\rm{w}_{adaptive}}+1}\right]+1$. Obviously, increasing the value of ${\rm w_{adaptive}}$ and the order of fitting polynomial cause almost fluctuations to be discarded, consequently it may suppress information of the underlying data sets. Fig. \ref{adap} indicates a schematic of partitioning in the adaptive detrending algorithm. Once the trend function for each partition to be determined, the corresponding adaptive detrended data in each segment is given by $x_i=z_i-\mathcal{Y}_{\nu}(i)$.\\
(II): Profile sets for both adaptive detrending series are  defined by:
\begin{eqnarray}
X(k) &\equiv& \sum_{i=1}^k \left[ x_i - \langle x \rangle \right]
\qquad k=1,\ldots,t \nonumber\\
Y(k) &\equiv& \sum_{i=1}^k \left[ y_i - \langle y \rangle \right]
\qquad k=1,\ldots,t \label{profile}
\end{eqnarray}
Since usually mean value has no considerable role in the final
results, and because we are going
to compare two different time series, therefore, we construct
data sets with zero mean and unit variance.\\

(III): We divide each mentioned profile into $N_s \equiv {\rm int}(t/s)$
non-overlapping segments of equal lengths, $s$, and for each segment
the fluctuation function is computed. To prevent the leakage of data when the size of the data sets is not a
multiple of scale, $s$, the same approach is done
from the opposite end, consequently one finds $2N_s$ segments.
\begin{eqnarray}
&&{\mathcal{F}}_{xy}(s,m) ={1 \over s} \sum_{i=1}^{s}\{X[(m-1) s+ i] - X_{\rm fit}(i,m)\}
\nonumber\\&&\qquad\qquad\qquad\qquad\times\{Y[(m-1) s + i] -
Y_{\rm fit}(i,m)\} \nonumber\\
\label{fsdef1}
\end{eqnarray}
for $m=1,...,N_s$ and:
\begin{eqnarray}
&&{\mathcal{F}}_{xy}(s,m) ={1 \over s} \sum_{i=1}^{s}\{X[t-(m-1) s+ i] - X_{\rm fit}(i,m)\}
\nonumber\\&&\qquad\qquad\qquad\qquad\times\{Y[t-(m-1) s + i] -
Y_{\rm fit}(i,m)\} \nonumber\\
\label{fsdef2}
\end{eqnarray}
for $m=N_s+1,...,2N_s$, where $X_{\rm fit}(i,m)$ and $Y_{\rm fit}(i,m))$ are arbitrary
fitting polynomials in $m$th segment.

Previous studies confirmed that common trends are eliminated by selecting linear fitting function. No trend means one should take
a zeroth-order fitting function \cite{PRL00}.\\

 (IV): On each local fluctuation function over all parts, the average is defined
by:
\begin{equation}
{\mathcal{F}}_{xy}(q;s) =\left\{ {1 \over  N_s} \sum_{m=1}^{ N_s} \left| {\mathcal{F}}_{xy}(s,m)
\right|^{q/2} \right\}^{1/q} \label{fdef}
\end{equation}
In principle, $q$ can take any real value, except zero. For $q=0$,
equation (\ref{fdef}) becomes:
\begin{equation}
{\mathcal{F}}_{xy}(0;s)= \exp\left( {1 \over  2N_s} \sum_{m=1}^{ N_s}\ln |{\mathcal{F}}_{xy}(s,m)|\right
) \label{fdef0}
\end{equation}
For $q=2$, the standard DCCA is retrieved.\\

(V): Finally, we demand that fluctuation functions behaves as power-law function and  the slope of the log-log plot
of ${\mathcal{F}}_{xy}(q;s)$ versus $s$ is determined as:
\begin{equation}\label{lam}
{\mathcal{F}}_{xy}(q;s) \sim s^{h_{^{xy}}(q)}
\end{equation}
If both underlying series are equal, $x=y$, then $ h_{xx}(q)=h_{xy}(q)$ and is nothing
else except the so-called generalized Hurst exponent, $h(q)$.
The  Hurst exponent ($0<H<1$) for non-stationary series is given by \cite{taqqu} (see the appendix of \cite{sadeghsun,sadeghriver} for more details) \begin{equation}\label{hurst11}
 H\equiv h_{xx}(q=2)-1.
 \end{equation}
The standard multifractal formalism shows that multifractal scaling exponent is \cite{bun02,zhi11}
\begin{equation}\label{tauqq}
\tau_{xy}(q)=qh_{xy}(q)-E.
\end{equation}
where $E$ is the fractal dimension of geometric support which is
$E=1$ for 1-Dimensional data set \cite{bun02}. The generalized
singularity spectrum, $f_{xy}(\alpha_{xy})$, of data is given by
so-called Legendre transformation of multifractal scaling exponent
as $f_{xy}(\alpha_{xy})=q\alpha_{xy}-\tau_{xy}(q)$ where
$\alpha_{xy}=\frac{\partial \tau_{xy}(q)}{\partial q}$ which is
known as H$\ddot{\rm o}$lder exponent.  For a multifractal series
$\alpha_{xy}$ has a spectrum instead of single value. The interval
of H$\ddot{\rm o}$lder spectrum, $\alpha_{xy}\in [\alpha^{\rm
min}_{xy},\alpha^{\rm max}_{xy}]$, can be determined by
\cite{muzy94,muzy95,halsey86}
\begin{eqnarray}\label{holder1}
\alpha^{\rm min}_{xy}&=&\lim_{q \rightarrow +\infty} \frac{\partial \tau_{xy}(q)}{\partial q},
\end{eqnarray}
\begin{eqnarray}\label{holder2}
\alpha^{\rm max}_{xy}&=&\lim_{q \rightarrow -\infty} \frac{\partial \tau_{xy}(q)}{\partial q}.
\end{eqnarray}
Up on the value of generalized Hurst exponent, $h_{xy}(q)$ is determined the correlation and power spectrum scaling exponents are determined. The correlation function for non-stationary process reads as:
\begin{eqnarray}
C_{xy}(t_1,t_2)&\equiv&\langle x(t_1)y(t_2)\rangle\nonumber\\
&\sim& [t_1^{-\gamma_{xy}}+t_2^{-\gamma_{xy}}-|t_1-t_2|^{-\gamma_{xy}}]
\end{eqnarray}
and $\gamma_{xy}=-2H_{xy}=2-2h_{xy}(q=2)$. The exponent of power
spectrum is also given by $\beta_{xy}=2H_{xy}+1=2h_{xy}(q=2)-1$. In
Table \ref{exponents}, we summarized the most relevant scaling
exponents for stochastic processes.
\begin{table}
\begin{center}
\medskip
\begin{tabular}{|c|c|c|c|c|}
  \hline
   Exponent & 1D-fGn & 1D-fBm &2D-Cascade& 2D-fBm \\
\hline
$H_{xy}$ & $h_{xy}(q=2)$ & $h_{xy}(q=2)-1$ &$h_{xy}(q=2)$& $h_{xy}(q=2)-1$\\
\hline
$\gamma_{xy}$ & $2-2H_{xy}$ & $-2H_{xy}$ &$1-2H_{xy}$& $-1-2H_{xy}$\\
\hline
$\beta_{xy}$ & $2H_{xy}-1$ & $2H_{xy}+1$ &$2H_{xy}$& $2H_{xy}+2$\\\hline
    \end{tabular}
\caption{\label{exponents}Some scaling exponents regarding stochastic processes in 1D and 2D. It must point out that the value of Hurst exponent for 2D case is under debated e.g. see \cite{zhou13,hosseinabadi12}.}
\end{center}
\end{table}

There is no guarantee  to have unique scaling exponent, $h_{xy}(q)$
for each $q$'s in all underlying scales, $s$. In mentioned situation
we should notice to range scale that underlying data sets is
investigated. Therefore, there are more than one statistical
behavior for underlying data sets in all scaling range. In other
words,  there exists either short-range cross-correlation or not at
all any cross-correlation. The $q$-dependency of $h_{xy}$ states
that cross-fluctuations have multifractal nature.  It must point out
that the importance of investigation of trends is based on at least
two following purposes: first of all, the existence of some kinds of
trends such as sinusoidal trend embedded in data sets, causes to
have cross overs in data
sets \cite{kunhu,trend2,f-dfa2,golub,trend3,trend3-1,physa,cooly65,koscielny98}.
In order to minimizing the effect of mentioned trends and find more
reliable statistical inference, some additional detrending
procedures are used e.g. F-DFA \cite{f-dfa,f-dfa2,movahedplasma}, Singular
Value Decomposition filtering (SVD) \cite{golub,trend3,trend3-1},
Empirical mode decomposition \cite{wu07}  and adaptive detrending algorithm \cite{hu09}.  After removing
global and local trends, we
obtain the fluctuation exponent by applying the MF-DXA.\\

\begin{figure}
\begin{center}
\includegraphics[width=1\linewidth]{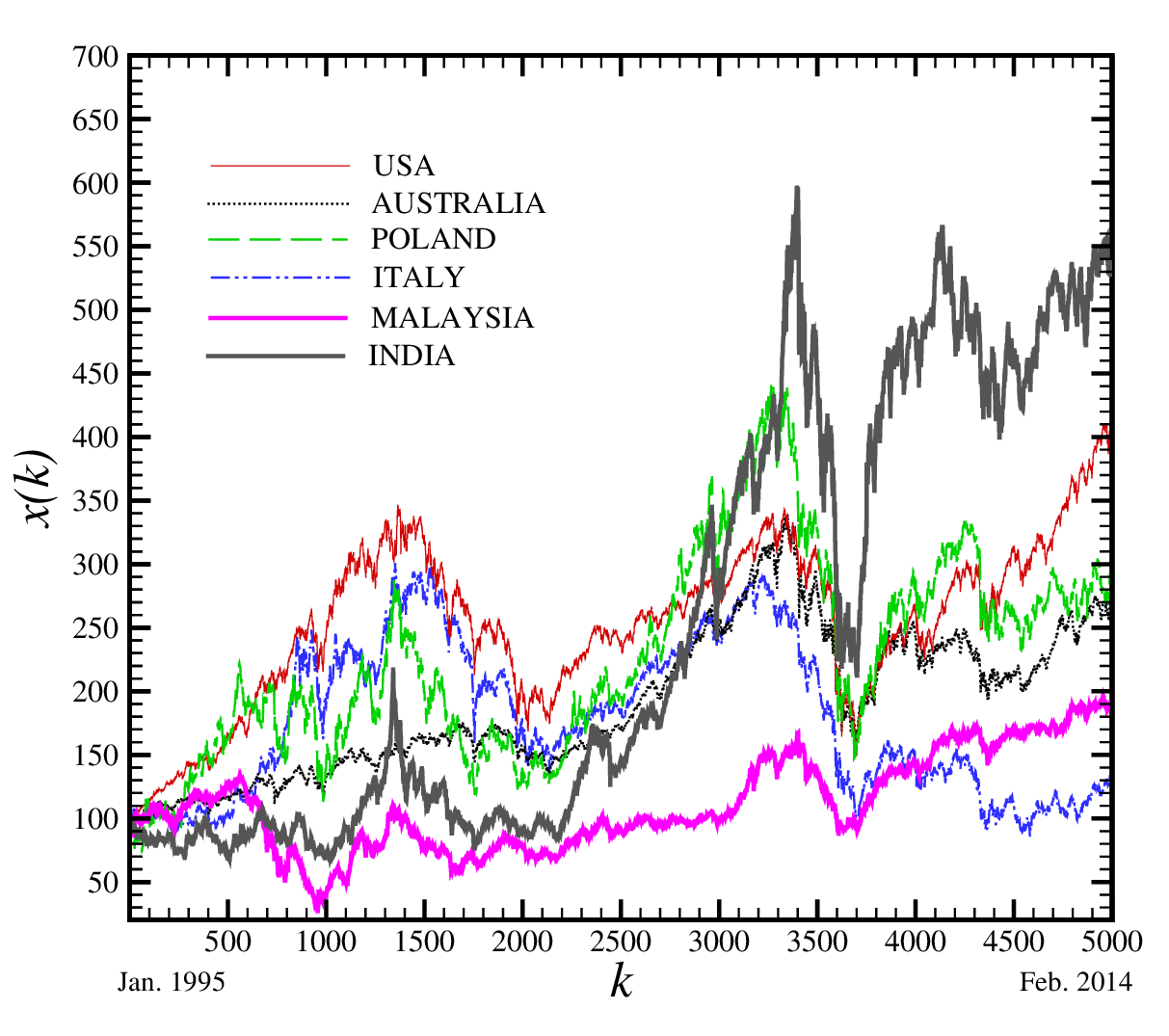}
\caption{\label{fig111} Time evolution of some monthly Stock market indices.}
\end{center}
\end{figure}

It has been demonstrated that to find the most reliable
value of scaling exponent based on DFA or DCCA methods, we should
set $s\leq (t/2)$, namely $N_s\ge 2$ \cite{bun02}.
To determine the slope of curve in the log-log plot of fluctuation
function versus scale (equation (\ref{lam})), we use likelihood
statistics as follows:
\begin{equation}
{\mathcal{L}}({\rm Data}|h_{xy}(q))\sim\exp\left(\frac{-\chi^2(h_{xy}(q))}{2}\right)
\end{equation}
 where:
\begin{equation}\chi^2(h_{xy}(q))=\sum_s\frac{[{\mathcal{F}}_{{\rm obs.}}(q;s)-{\mathcal{F}}_{{\rm fit}}(s;h_{xy}(q))]^2}{\sigma_{{\rm obs.}}^2(q;s)}\end{equation}
Here  ${\mathcal{F}}_{{\rm fit}}(s;h_{xy}(q))$ and ${\mathcal{F}}_{{\rm obs.}}(q;s)$ are
fluctuation functions determined by equation (\ref{lam}) and computed directly from the data set by using
DFA or DCCA, respectively. Also,
$\sigma_{{\rm obs.}}(q;s)$ is the mean standard deviation, associated
with ${\mathcal{F}}_{\rm obs.}(q;s)$. Maximizing likelihood function corresponds to minimizing $\chi^2$ for best value of $h_{xy}(q)$. The value of error-bar at
$1\sigma$ confidence interval of $h_{xy}(q)$ is computed by the
likelihood function based on the following condition:
\begin{equation}
68.3\%=\int_{-\sigma^{-}(q)}^{+\sigma^{+}(q)}{\mathcal{L}}({\rm Data}|h_{xy}(q))dh_{xy}(q)
\end{equation}
The best fit value of scaling exponent at $1\sigma$
confidence interval will be reported according to $h_{xy}(q)_{-\sigma^-(q)}^{+\sigma^+(q)}$
for each moment, $q$'s. In the Gaussian case apparently, $\sigma^-(q)=\sigma^+(q)$.

To make our results more sense and complete, we follow approach introduced by G.F. Zebende \cite{zeb11} for so-called cross-correlation coefficient as:
\begin{equation}\label{dcca1}
\sigma_{\rm DCCA}(s)\equiv \frac{{\mathcal{F}}_{xy}^2(s)}{{\mathcal{F}}_{xx}(s){\mathcal{F}}_{yy}(s)}
\end{equation}
where
\begin{equation}
{\mathcal{F}}_{xy}^2(s)\equiv\frac{1}{N_s}\sum_{m=1}^{N_s}{\mathcal{F}}_{xy}(s,m)
\end{equation}
and ${\mathcal{F}}_{xy}(s,m)$ is given by Eqs. (\ref{fsdef1}) and (\ref{fsdef2}). Finally we compute $\sigma_{\rm DCCA}=\langle \sigma_{\rm DCCA}(s)\rangle_s$. The $\sigma_{DCCA}=+1$ corresponds to prefect cross-correlation,  $\sigma_{DCCA}=0$  indicates no cross-correlation between underlying data sets while $\sigma_{DCCA}=-1$ demonstrates completely anti-cross correlation. Meanwhile, a modified version for quantifying cross-correlation based on Eq. (\ref{dcca1}) namely $q$-dependent cross-correlation coefficient has been introduced by Kwapie\'{n} et. al, \cite{dccaq15},  we confine ourselves to use mentioned measure just for $q=2$ and in another study, we will take into account for more values of $q$.

\begin{figure}
\begin{center}
\includegraphics[width=1\linewidth]{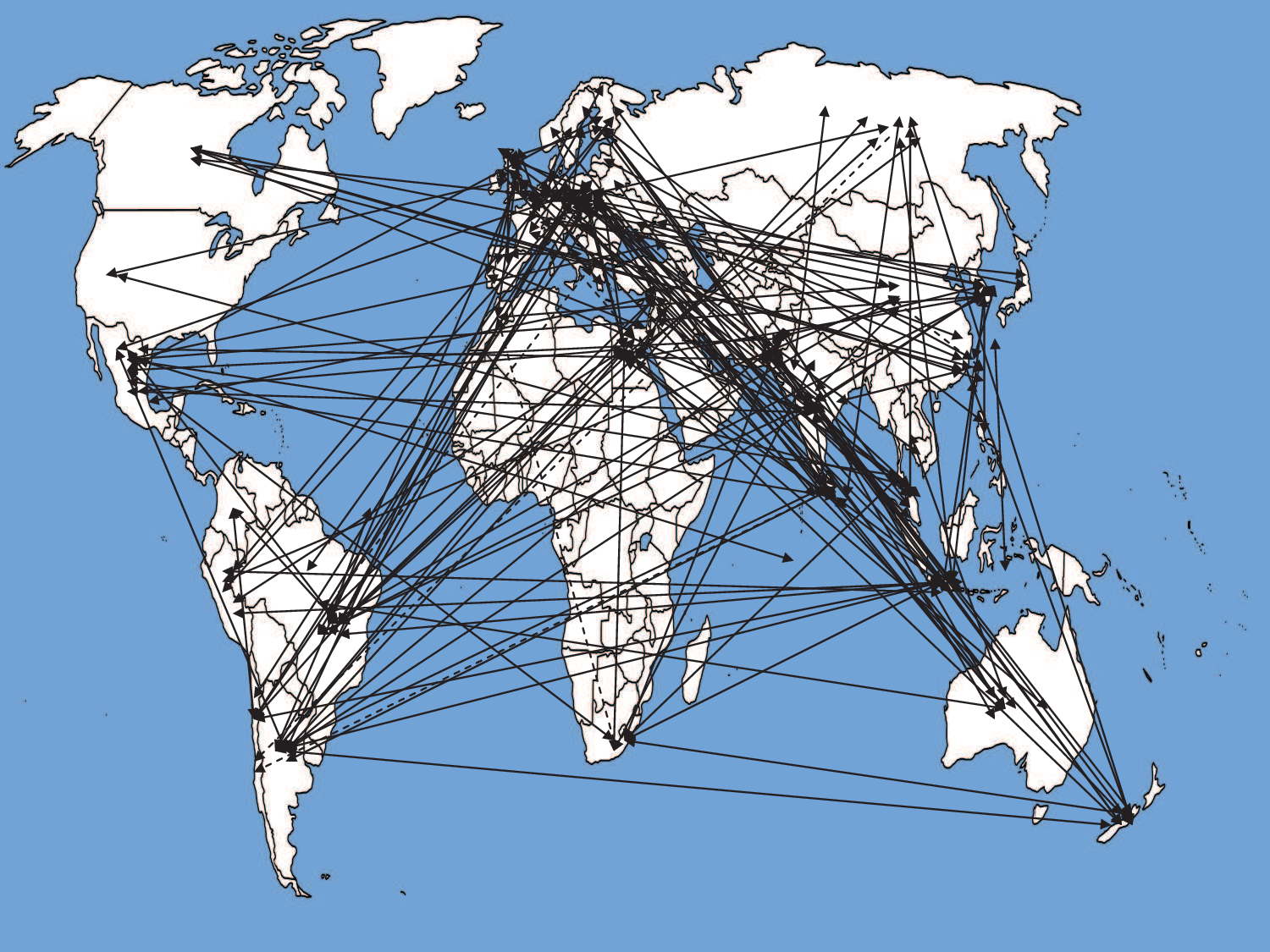}
\caption{\label{fig112} Long-range significant relationships between Stock markets (dashed-line indicates bilateral relationships, and solid line indicates unilateral relationships).}
\end{center}
\end{figure}

\begin{figure}
\begin{center}
\includegraphics[width=1\linewidth]{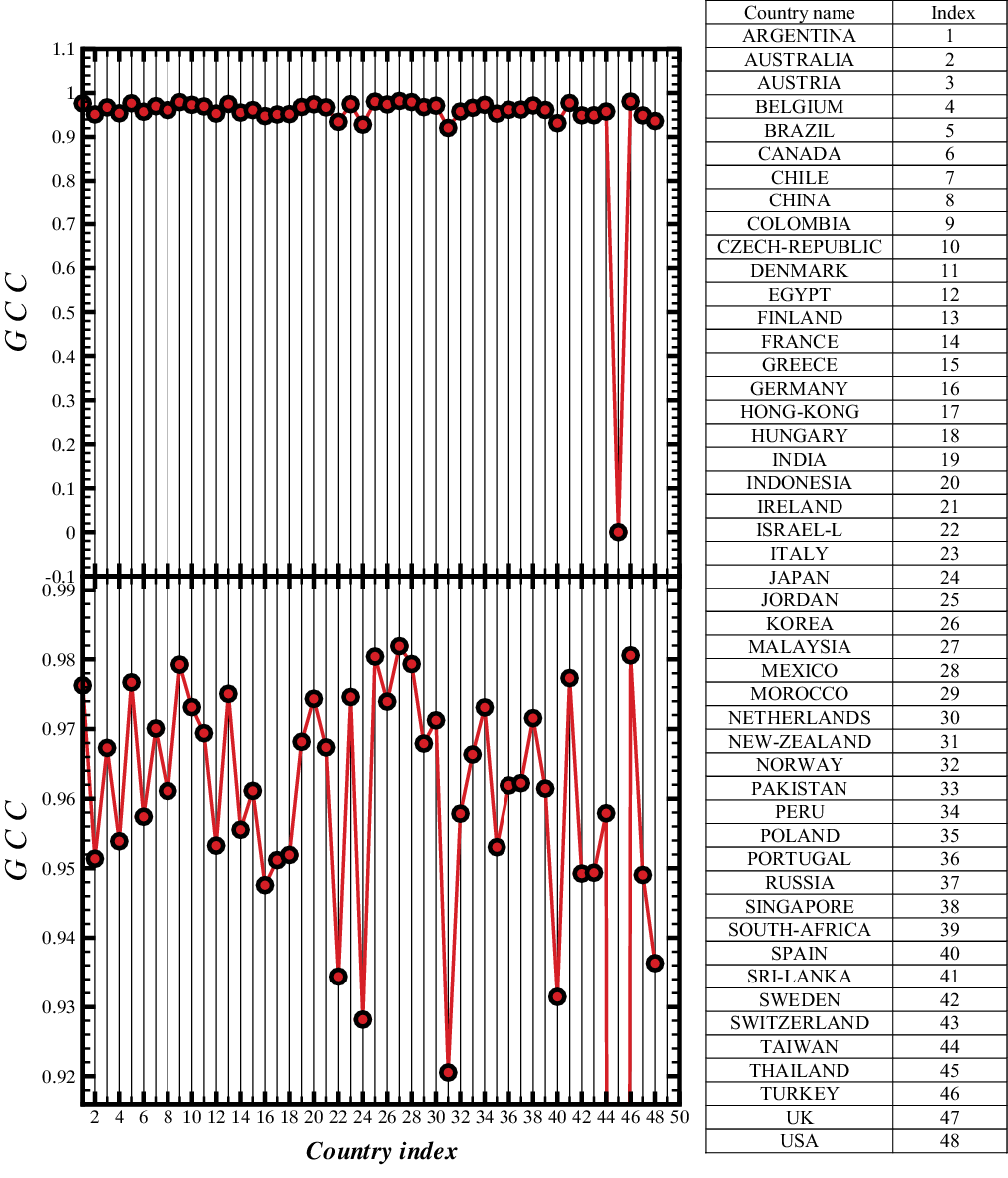}
\caption{\label{fig113} Top panel: Average global correlation coefficient (GCC) for each country. Bottom panel: Zoom of the average global correlation for each country.}
\end{center}
\end{figure}

\section{Data description}
The data used in this paper consists on adjusted market
capitalization Stock market indices of 48 developed and emerging
markets, constructed by Morgan Stanley Capital International (MSCI)
and downloaded from DataStream. We use daily index prices over the
period January 1995 to February 2014, corresponding to 4995
observations per index. The MSCI classification depends on three
criteria: economic development, size and liquidity and market
accessibility and divide markets on developed, emerging and frontier
markets (for more details, see http://www.msci.com). Our database
includes 23 markets classified as developed, 21 markets classified
as emerging and 4 frontier markets. The developed markets are:
Canada, United States (from America), Austria, Belgium, Denmark,
Finland, France, Germany, Ireland, Israel, Italy, the Netherlands,
Norway, Portugal, Singapore, Spain, Sweden, Switzerland, United
Kingdom (from Europe), Australia, Honk Kong, Japan, New Zealand and
Singapore (from the Pacific). The emerging markets are Brazil,
Chile, Colombia, Mexico, Peru (from Americas), the Czech Republic,
Egypt, Greece, Hungary, Poland, Russia, South Africa, Turkey
(Europe, Middle East \& Africa), China, India, Indonesia, Korea,
Malaysia, Philippines, Taiwan, and Thailand (Asia). The frontier
markets are Argentina (Americas), Morocco (Africa), Jordan (Middle
East) and Pakistan (Asia). The data are the relative price indices
for these markets, where the base 100 was set in the first
observation. In order to illustrate the behavior of those markets,
we present the time evolution of some markets in Fig. \ref{fig111}.
In a very simplistic way, we can observe some similar behavior
between some Stock markets, besides the differences of scale. For
example, in the developed markets group, the Europe shows some
"synchronization", such as some markets of Asia, namely Singapore
and Japan. The emerging markets also seem to show high levels of
"synchronization" or similar behavior, especially in Europe and
South America. It is important to note the higher values of the
Turkish Stock market, which may induce that this market had strong
increment on the period under analysis. Of course, this kind of
analysis is merely preliminary. In order to evaluate the relations
between those markets, it is important to use robust techniques in
linear and nonlinear terms.

\section{Results of Stock market comovements}
In the previous section the mathematical and computational tools to extract reliable information regarding the underlying data have been explained. In this section we are going to apply mentioned methods on the series.

\subsection{Evidence of cointegration and causality tests}
We tested the stability of our time series by regressing it on a
nonsignificant constant. The results indicate the presence of
structural breaks for all the variables and some of those structural
breaks seems to be related with the existence of Stock market
crashes and financial crisis. Given that all variables are
non-stationary, we considered the possibility of estimating a
long-run relationship between all these variables. To test for
cointegration between all those series, we used the Phillips tests
suggested by Gregory and Hansen \cite{greg96}  because the power of
the Johansen's test may be reduced substantially when the series
exhibits structural breaks. Using the Gauss code provided by Bruce
Hansen, we tested the presence of cointegration in all pairs of
variables (more precisely, the respective logarithms) and our
results indicate the existence of 170 bivariate cointegration
vectors. These long-range relationships were evaluated in terms of
linear causality, using the VECM (Vector Error Correction Model)
since the Granger causality test can not be used for nonstationarity
variables (the number of lags were selected using AIC and BIC). Fig.
\ref{fig112} shows the long-range significant relations obtained
using the VECM and respective type of relation (bilateral and
unitarily).

Taking only Fig. \ref{fig112} as a reference, it is extremely
difficult to analyze all the significant relations. Although, this
figure helps the reader to understand five main facts: (i) Brazil,
Colombia, Egypt, India, Indonesia, Austria and Australia are the
Stock markets that show cointegration with more foreign markets;
(ii) Granger causality allows to analyze the direction of those
relations and significant number of bi-directional relationships is
greater than the number of unidirectional relations; (iii) Relations
between emerging markets seems to be more pronounced; (iv) USA does
not seem to be the motor; (v) Strong relationships within Europe and
between Europe and South America.

\begin{widetext}

\begin{table}
\begin{center}
\medskip
\begin{tabular}{|c|c|c|c|c|c|}
  \hline
    Criterion name & $dd$ & $de$ & $ee$ & $ef$ & $ff$  \\\hline
  Mean & $0.959067$ &  $ 0.948624$ & $0.969767 $ & $0.977078$ & $0969962$  \\\hline

  Median & $0.96941$ &  $ 0.96059$ & $0.97738 $ & $0.97887$ & $0.9771198$ \\\hline
  Standard-deviation & $0.027804$ &  $0.036165$ & $0.02653 $ & $0.007469$ & $0.022081$ \\\hline
  Kurtosis & $4.820176$ &  $ 1.097552$ & $7.553805$ & $1.074349$ & $3.155796$ \\\hline
  Asymmetry & $-2.200398$ &  $-1.278313$ & $-2.712298 $ & $-0.995404$ & $-1.680047$ \\\hline
Minimum & $0.86383$ &  $0.84784$ & $0.870824 $ & $0.95692$ & $0.93771$\\\hline
  Maximum & $0.98162$ &  $ 0.992808$ & $0.99287 $ & $0.98999$ & $0.98774$ \\\hline
  $N$ & $27$ &  $ 65$ & $47 $ & $27$ & $4$ \\\hline
\end{tabular}
\caption{\label{tab1} Descriptive statistics for the GCC. Note that $dd$ refers to the GCC of \underline{d}eveloped versus \underline{d}eveloped indices, $de$ corresponds to \underline{d}eveloped versus \underline{e}merging, $ee$ represents \underline{e}merging versus \underline{e}merging, $ef$ shows \underline{e}merging versus \underline{f}rontier and $ff$ refers to \underline{f}rontier versus \underline{f}rontier indices.}
\end{center}
\end{table}
\end{widetext}

\subsection{Mutual information}
Robust methods are needed to have reliable results from statistical analysis of various fluctuations recorded in the nature. In order to evaluate the long-range relations between Stock markets as a whole, we perform a similar analysis using mutual information independence test and the global correlation coefficient (GCC). Since mutual information may lose some properties in presence of non cointegrated and non stationary series, we estimate this measure only for the pairs of indices that show evidence of cointegration. Given this, we estimate the mutual information and the global correlation coefficient for 170 pair of indices and according to the relevant critical values \cite{dion06} all the obtained values are statistically significant. Fig. \ref{fig113} shows the average of the global correlation coefficient obtained for each country. As we may see from this figure, most of the countries exhibit high values (between 0.9 and 1) except Taiwan. This result is explained by the fact that Taiwan did not show evidence of cointegration with any other country, not allowing the estimation of MI.

If we analyze the part (b) of Fig. \ref{fig113}, we have the
possibility to differentiate in a better way the values of average
GCC for each country. It is worth to note that Colombia, Mexico and
Jordania seem to evidence higher values of the average of GCC. Would
this mean that these Stock indices are more related with the rest of
the world, than others? Another important aspect to
refer is the fact that these results do not match perfectly with the
results obtained with the linear approach. The main differences may
be related with the fact that mutual information has the ability to
capture linear and nonlinear dependence, without imposing any
assumptions in terms of structure and probability distribution.
Measures of information theory, namely entropy, metric entropy and
mutual information, show evidence of strong robustness in the
evaluation of serial and cross dependence between vectors of
variables \cite{darb00,lin94,gran04}. In order to better understand the level of global
relations between the several types of indices, we estimate some
descriptive statistics (Table \ref{tab1}).  Note that $dd$ refers to the GCC of \underline{d}eveloped versus \underline{d}eveloped indices, $de$ corresponds to  \underline{d}eveloped versus \underline{e}merging, $ee$ shows \underline{e}merging versus \underline{e}merging, $ef$ represents \underline{e}merging versus \underline{f}rontier and $ff$ refers to \underline{f}rontier versus \underline{f}rontier indices. Table \ref{tab1} shows that the highest level of serial correlations between Stock indices occur between emerging and frontier Stock indices and between emerging Stock indices. The statistical analysis of GCC indicates, for the negative asymmetry and also leptokurtosis a stronger concentration of coefficients around the mean values. Being those mean values very close to 1, we should conclude that the highest levels of global correlations (in terms of concentration around mean) are between indices from the same type of stock market ($dd$, $ee$ and $ff$). Of course this comparison is not enough to make conclusions. Given this, we performed the ANOVA test, which results point to the rejection of the null hypothesis ($H0 :\mu_{dd}=\mu_{ee} =\mu_{ef} =\mu_{ff}$). In order to compare all the means involved in this study we also performed some Scheff$\acute{e}$ tests and results show that there are significant differences between the mean of GCC of some groups. For example, the mean of GCC of developed versus emerging is smaller that the respective means of emerging versus emerging stock markets. A similar conclusion can be taken related to emerging versus frontier stock markets. This may indicate that the relations between developed stock markets are, probably, not so pronounced which can be explained by the maturity of those markets and possible high levels of efficiency.

\begin{figure}
\begin{center}
\includegraphics[width=0.875\linewidth]{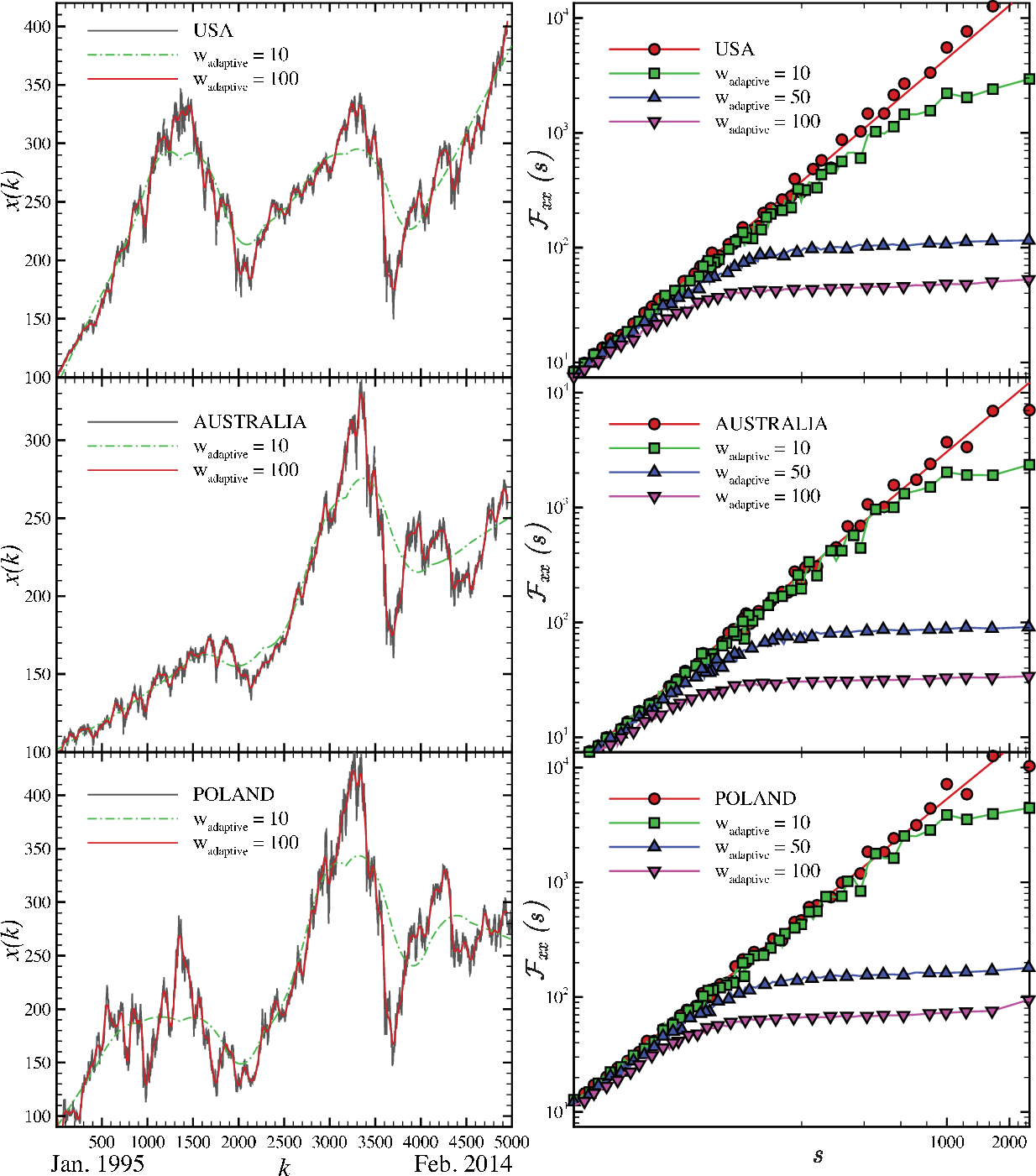}
\includegraphics[width=0.87\linewidth]{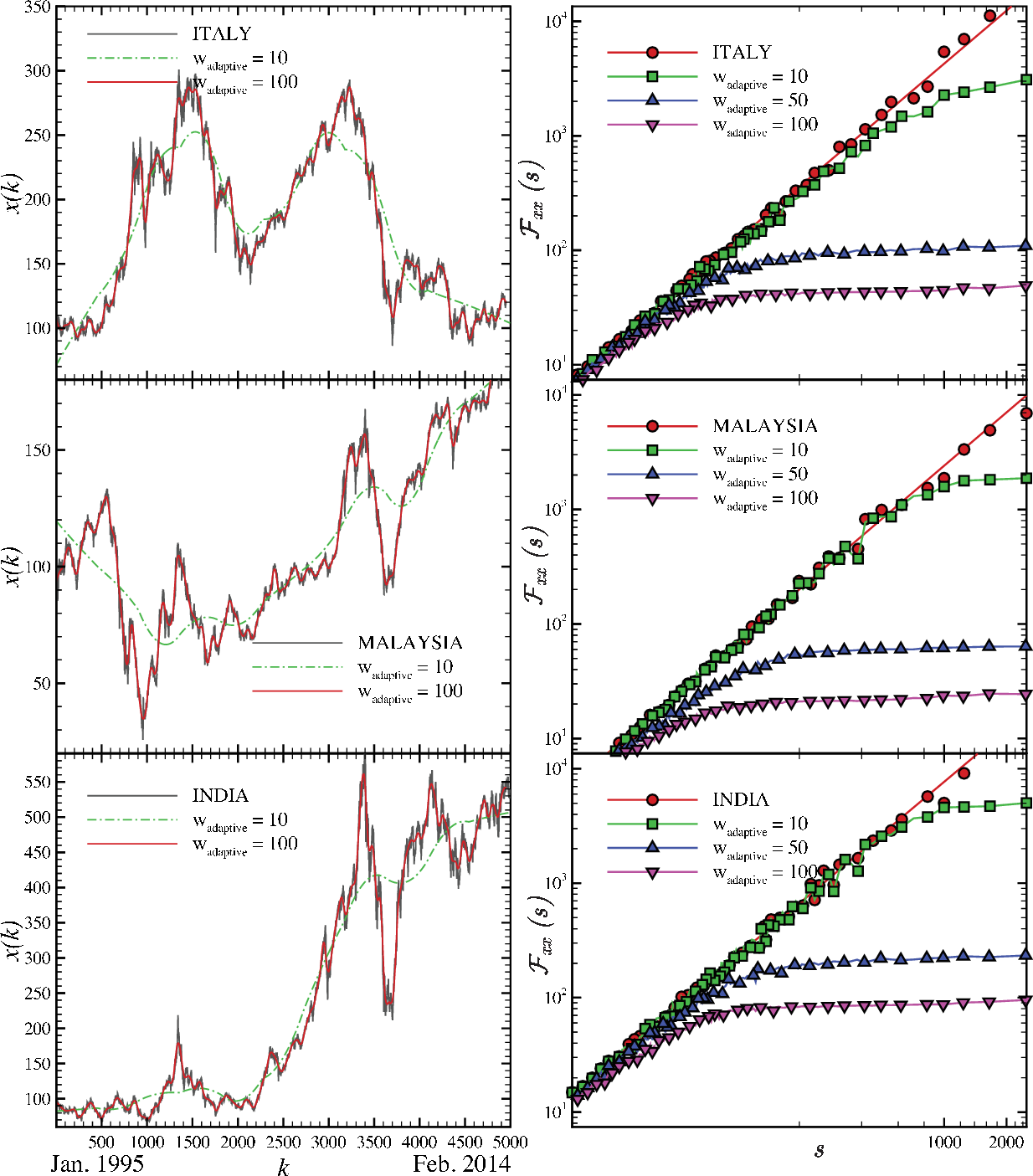}
\caption{\label{figfs1} Left panels show data and trend functions computed for ${\rm w_{adaptive}}=10$ (dash-dot line) and ${\rm w_{adaptive}}=100$ (solid line). Right panels correspond to fluctuation function for some typical data sets used in this paper as a function of scale. Circle symbol indicates ${\mathcal{F}}_{xx}(s)$ versus $s$ for the regular DFA, while other symbols show the results  for adaptive detrending approach.}
\end{center}
\end{figure}

\begin{figure}
\begin{center}
\includegraphics[width=1\linewidth]{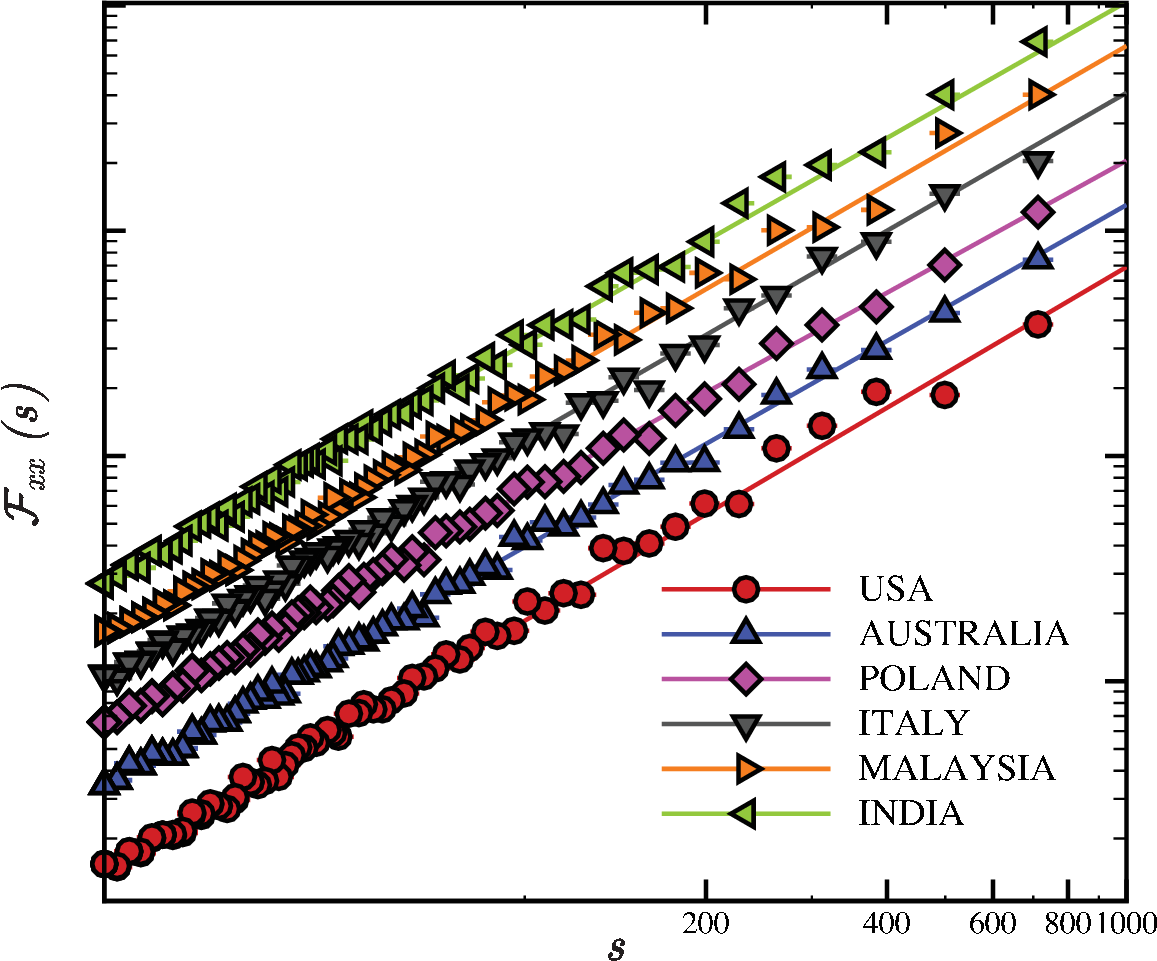}
\caption{\label{figfs} Fluctuation function for some typical data sets used in this paper as a function of scale. To make more sense we shifted the value of ${\mathcal{F}}_{xx}(s)$ vertically,}
\end{center}
\end{figure}

\begin{table}
\begin{center}
\scalebox{0.6}{
\medskip
\begin{tabular}{|c|c|c|c|}
  \hline
    Index & Hurst & $\gamma_{xx}$&$\Delta \alpha_{xx}$ \\\hline
   1&    $ 0.571  \pm  0.007$ & $ -1.142  \pm  0.013 $ & $ 0.380\pm 0.509 $ \\\hline
   2 & $ 0.518\pm0.005$&$ -1.036 \pm0.009 $& $ 0.304\pm  0.435 $\\\hline
   3&$ 0.601  \pm  0.006$&$  -1.202 \pm  0.012$&$ 0.421 \pm  0.678  $\\\hline
   4 &$ 0.566\pm  0.005$&$  -1.133   \pm 0.009$ & $ 0.544\pm       0.683$\\\hline
    5 &$   0.502\pm    0.005$&$   -1.004    \pm   0.011$&$  0.620\pm 0.617$\\\hline
    6 &$ 0.529\pm 0.005$&$    -1.059     \pm    0.011$&$0.833\pm  0.586$\\\hline
  7&$   0.498  \pm   0.004$&$   -0.996     \pm 0.008 $&$ 0.496\pm 0.486 $\\\hline
 8 &$   0.485  \pm     0.005 $&$  -0.971    \pm  0.009$ &$   0.579\pm 0.666 $\\\hline
 9&$  0.551   \pm 0.006$& $ -1.103\pm 0.011$&$  0.607\pm 0.522$ \\\hline
10&$ 0.520 \pm 0.006$&$  -1.041    \pm  0.013$&$  0.718\pm  0.715 $\\\hline
11&$    0.550  \pm   0.005$&$ -1.099    \pm 0.009$&$   0.822\pm 0.631$\\\hline
 12&$  0.518     \pm 0.019$&$   -1.037\pm  0.039$&$  0.905\pm1.783$\\\hline
 13&$ 0.526 \pm 0.006$&$  -1.052   \pm0.011$&$ 0.528\pm0.531$\\\hline
 14&$ 0.551 \pm 0.004$&$   -1.101     \pm   0.007$&$ 0.587\pm  0.573$\\\hline
15&$0.549\pm 0.005$&$ -1.098    \pm 0.010$&$0.682\pm 0.607$\\\hline
 16&$ 0.541   \pm 0.004$&$  -1.083   \pm 0.008$&$0.493\pm0.531$\\\hline

 17&$ 0.494     \pm 0.005$&$  -0.988     \pm 0.010$&$ 0.395\pm 0.475$\\\hline
 18&$ 0.509\pm 0.006$&$ -1.018   \pm 0.012$&$ 0.891\pm  0.696$\\\hline
 19&$   0.518  \pm 0.005$&$ -1.036     \pm 0.010$&$0.462\pm 0.658$\\\hline
 20&$  0.532    \pm 0.004$&$  -1.064     \pm 0.008$&$ 0.681\pm  0.487$\\\hline
 21&$    0.570   \pm 0.005 $&$  -1.140    \pm 0.010$&$  0.391\pm 0.507$\\\hline
 22&$  0.510\pm 0.004 $&$ -1.020   \pm   0.008$&$  0.439\pm  0.381$\\\hline
 23&$  0.540\pm 0.004$&$  -1.076    \pm  0.008$&$    0.373\pm 0.436$\\\hline
 24&$ 0.522 \pm 0.004$&$ -1.044     \pm0.008 $&$ 0.246\pm  0.473 $\\\hline
 25&$ 0.602  \pm   0.006$&$   -1.204    \pm  0.012$&$0.572\pm 0.471$\\\hline
 26&$ 0.475\pm 0.004$&$  -0.950    \pm 0.008$&$  0.461\pm0.423$\\\hline
 27&$  0.545\pm 0.004$&$  -1.090    \pm  0.008$&$ 0.477\pm 0.398$\\\hline
  28&$    0.528  \pm0.004$&$ -1.056    \pm0.008$&$ 0.512\pm 0.646$\\\hline
 29&$  0.571 \pm0.005$&$   -1.142    \pm\  0.010$&$ 0.839\pm 0.640$\\\hline
 30&$    0.538  \pm 0.004$&$   -1.076    \pm0.008$&$  0.716\pm 0.538$\\\hline
 31&$  0.457     \pm   0.004$&$  -0.914 \pm0.008$&$  0.467\pm 0.399$\\\hline
 32&$   0.523  \pm0.006$&$    -1.046 \pm  0.012 $&$  0.447\pm       0.539 $\\\hline
 33&$    0.476  \pm0.017 $&$  -0.952  \pm 0.034 $&$0.729\pm  1.688 $\\\hline
34&$ 0.515   \pm  0.005$&$    -1.030    \pm  0.010$&$   0.603\pm    0.530$\\\hline
 35&$    0.473 \pm   0.005$&$  -0.946 \pm   0.010$&$0.512\pm 0.471$\\\hline
 36&$  0.584 \pm0.004$&$    -1.168    \pm   0.008$&$ 0.469\pm     0.504$\\\hline
 37&$  0.536 \pm0.007$&$  -1.072     \pm0.014$&$ 0.492\pm 0.686$\\\hline
38&$   0.524\pm0.006$&$  -1.048     \pm 0.012 $&$  0.466\pm 0.487$\\\hline
 39&$   0.517  \pm0.004$&$   -1.034     \pm   0.008$&$   0.448\pm    0.502$\\\hline
 40&$   0.531\pm 0.004$&$    -1.062     \pm 0.008$&$ 0.806\pm    0.328$\\\hline
 41&$   0.562   \pm0.005 $&$ -1.124    \pm 0.010$&$  0.745\pm 0.742$\\\hline
42&$  0.557\pm    0.004$&$    -1.114     \pm   0.008$&$  0.537\pm0.429$\\\hline
 43&$    0.534  \pm  0.004$&$   -1.068     \pm0.008$&$   0.561\pm      0.395$\\\hline
 44&$ 0.473\pm0.005$&$  -0.946     \pm0.010$&$   0.367\pm  0.439$\\\hline
45&$   0.515        \pm0.005  $&$   -1.030 \pm0.010$&$  0.424\pm 0.493$\\\hline
 46&$    0.500     \pm    0.005$&$   -1.000 \pm   0.010$&$ 0.868\pm0.806$\\\hline
 47&$  0.472   \pm0.005$&$ -0.944     \pm0.010$&$  0.534\pm  0.483$\\\hline
 48&$   0.531  \pm0.005$&$   -1.062     \pm 0.010$&$0.819\pm  0.457$\\\hline
\end{tabular}}
\caption{\label{index}The name of countries, their indices, Hurst ($H\equiv h_{xx}(q=2)-1$) and correlation, $\gamma_{xx}$,  exponents.The value of $\Delta \alpha_{xx}=\alpha_{xx}^{\rm max}-\alpha_{xx}^{\rm min}$ given by equations (\ref{holder1}) and (\ref{holder2}) just it has been done for same data.}
\end{center}
\end{table}

\begin{figure}
\begin{center}
\includegraphics[width=0.7\linewidth]{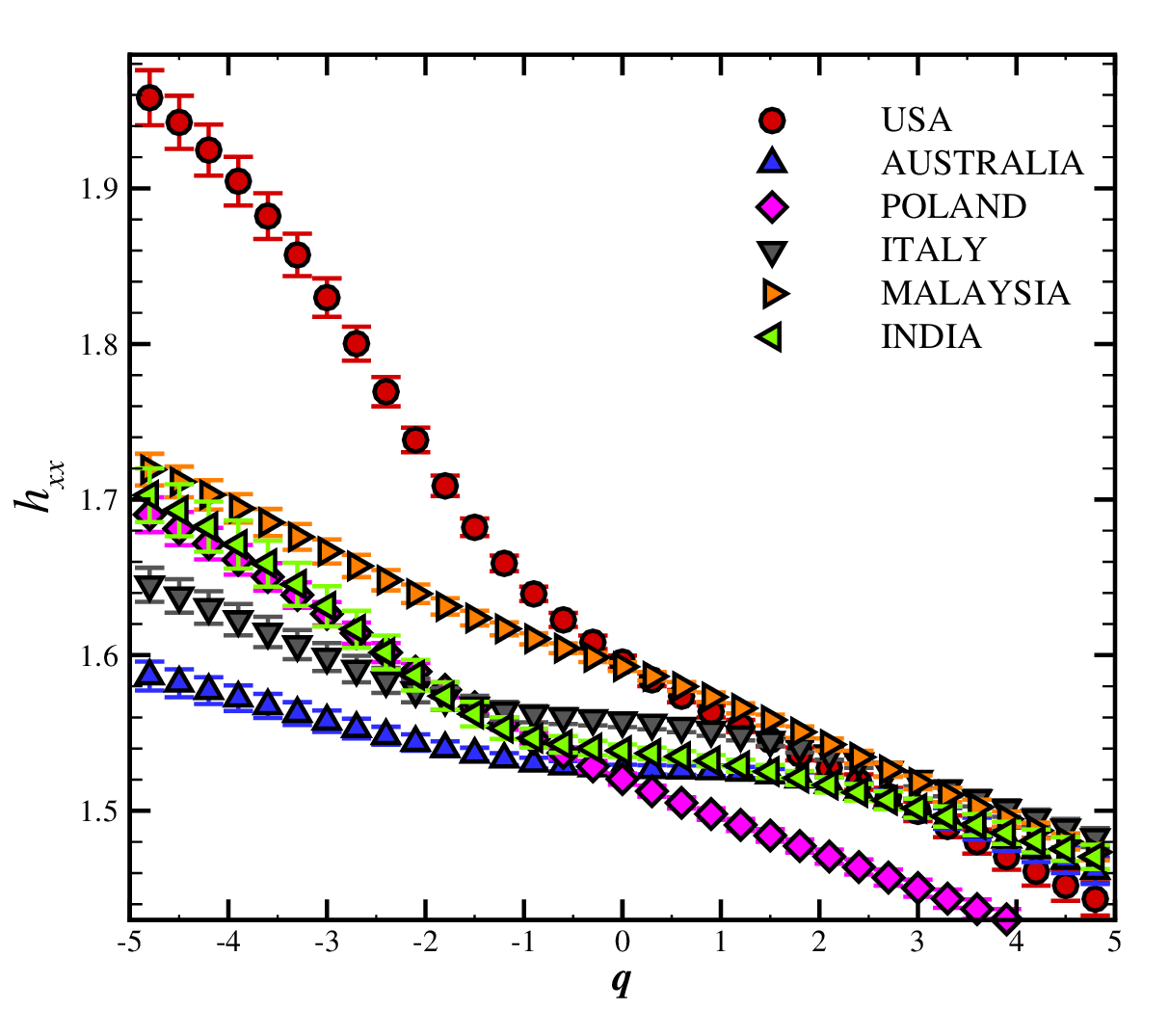}
\includegraphics[width=0.7\linewidth]{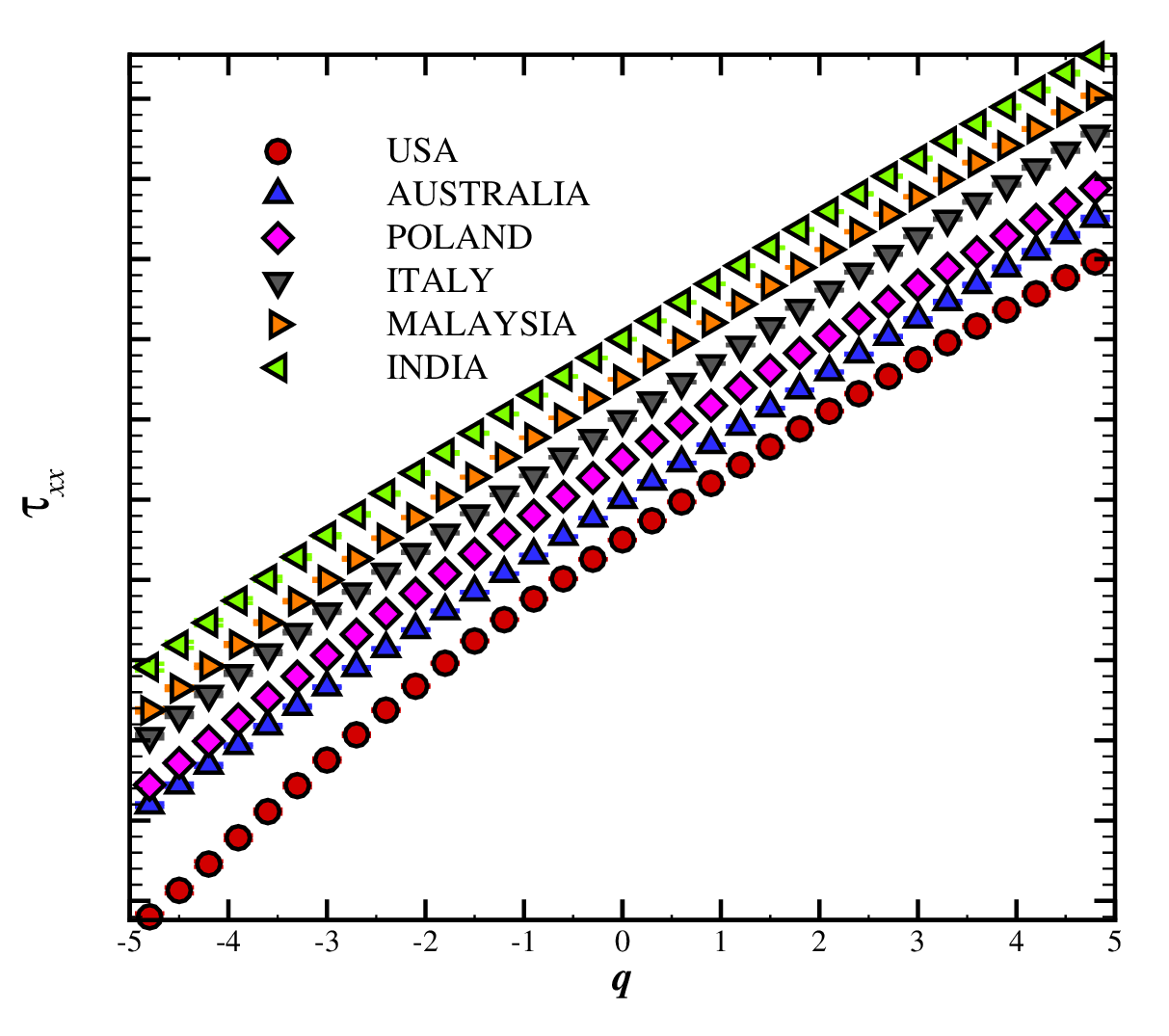}
\includegraphics[width=0.7\linewidth]{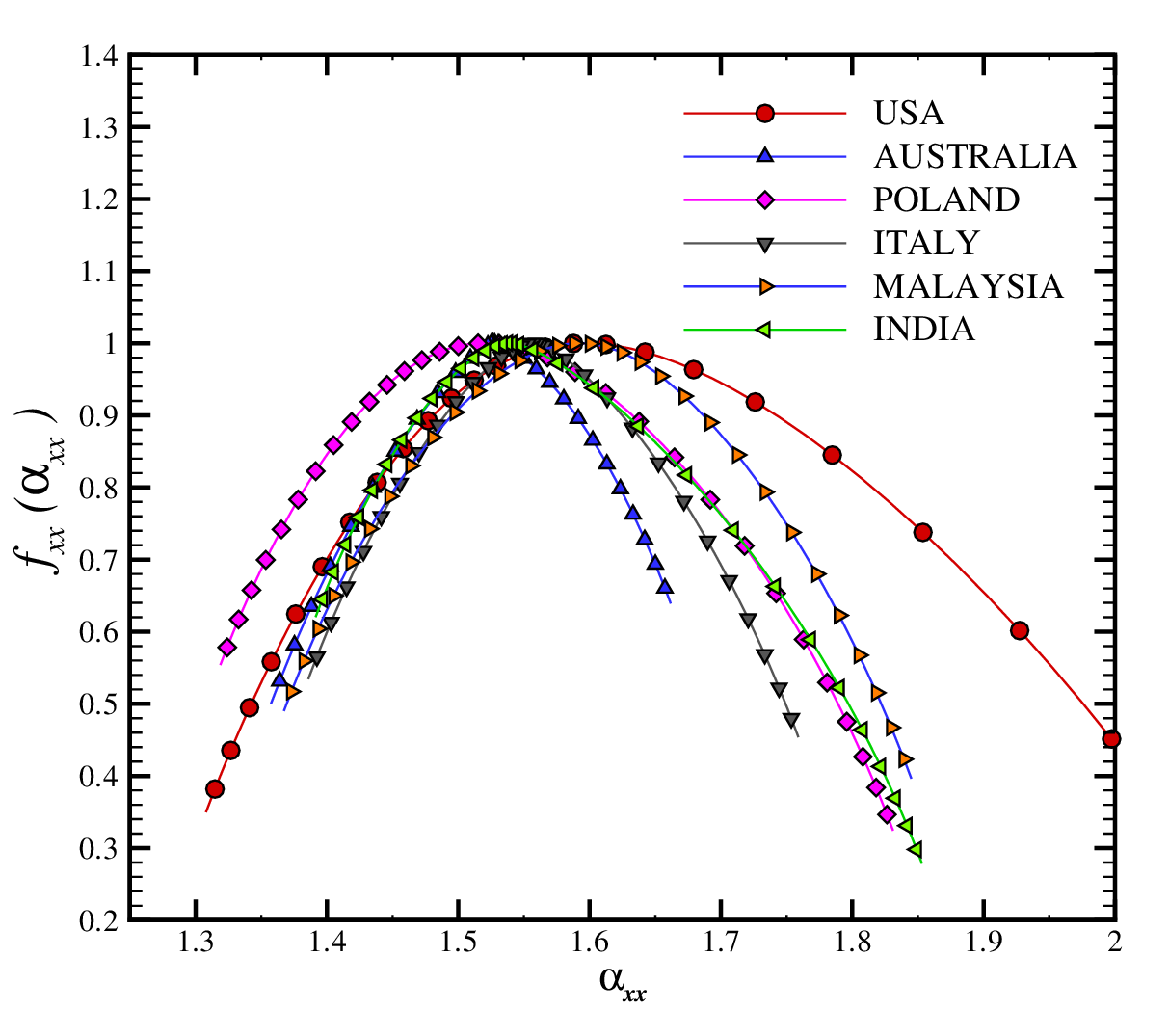}
\caption{\label{fighq} Generalized Hurst exponent and multifractal exponent of some Stock markets as a function of $q$. For $\tau_{xx}(q)$ we shifted the value through vertical axis. Lower panel shows singularity spectrum for some of data sets.}
\end{center}
\end{figure}

\begin{figure}
\begin{center}
\includegraphics[width=0.7\linewidth]{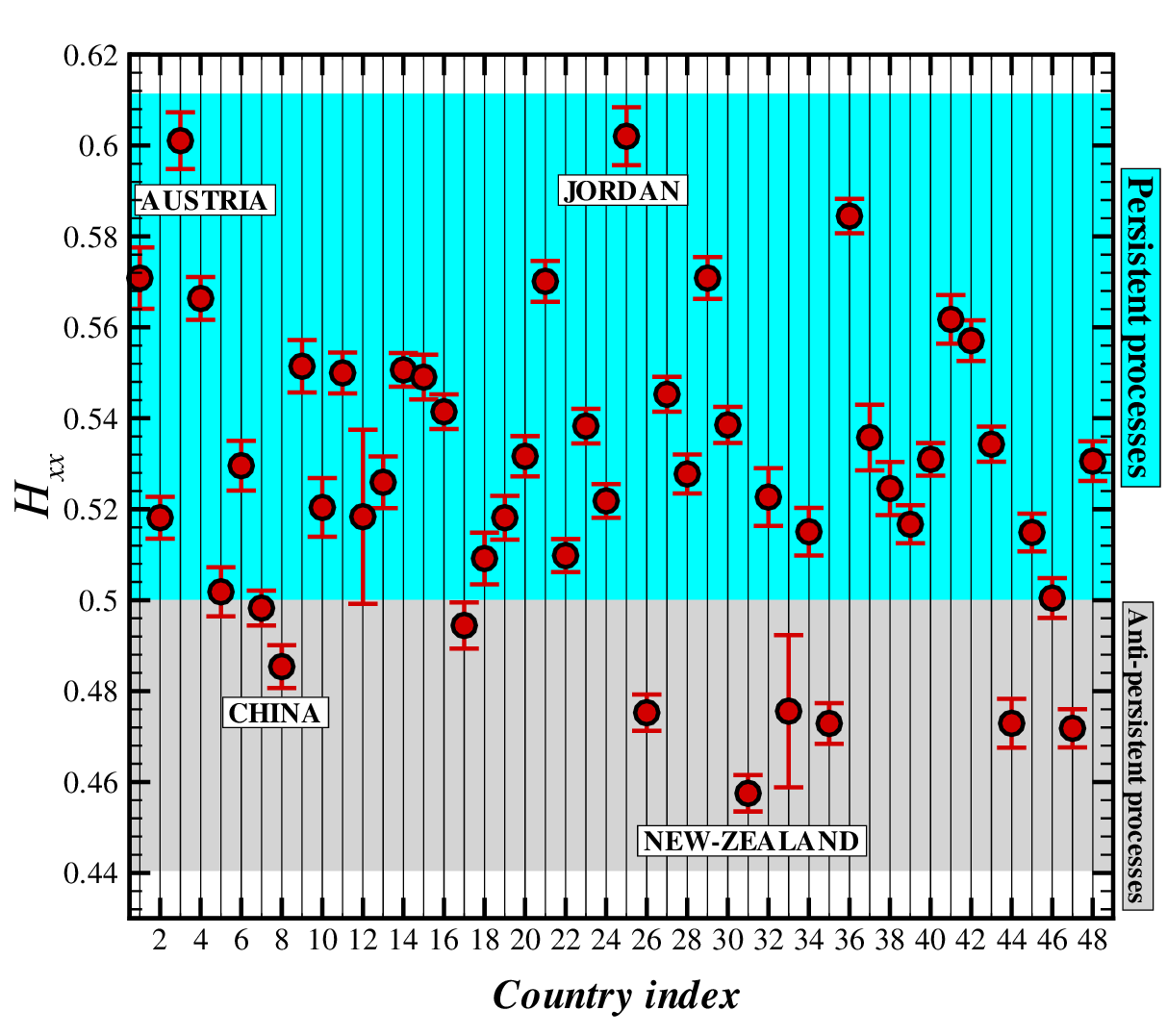}
\includegraphics[width=0.7\linewidth]{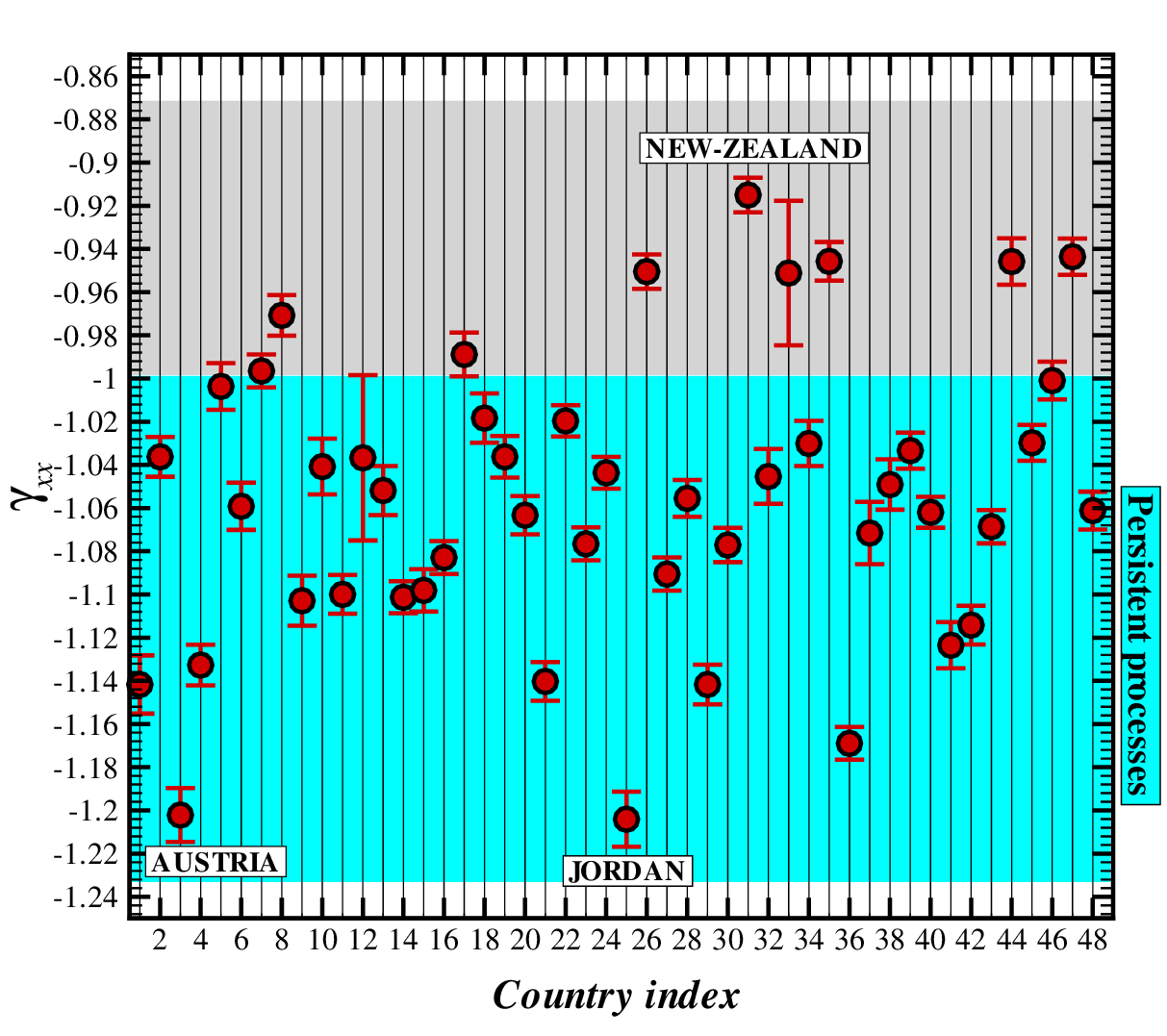}
\includegraphics[width=0.7\linewidth]{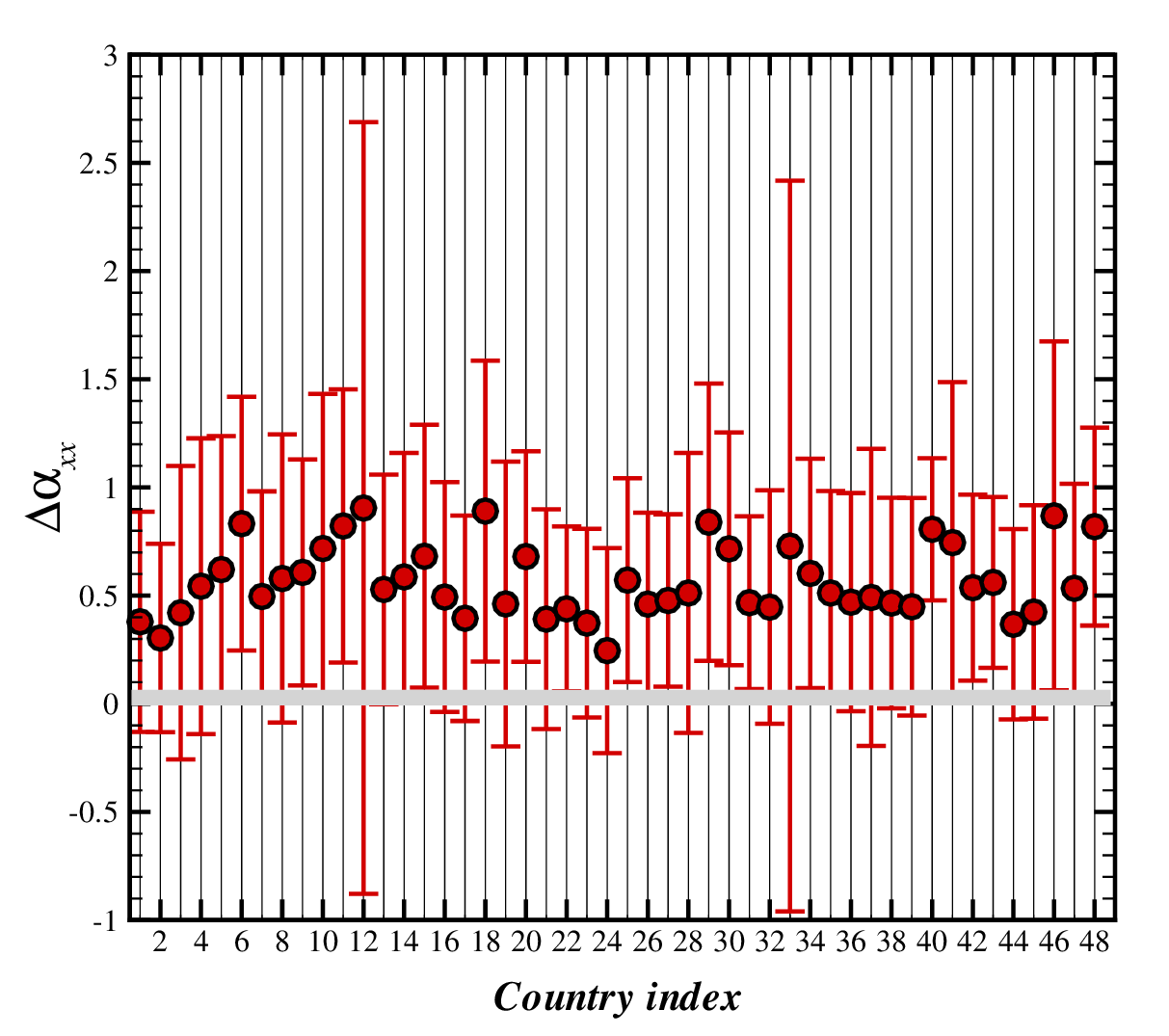}
\caption{\label{fig1} Upper panel corresponds to the value of Hurst exponent, $H_{xx}$ given by DFA for 48 countries used in this paper (see the text). Lower panel shows $\gamma_{xx}$ for same data. The width of H$\ddot{\rm o}$lder exponent corresponding multifractality nature of series has been indicated in lower panel.}
\end{center}
\end{figure}

\subsection{Implementation of AMF-DFA and AMF-DXA to Stock indices}

In this subsection we report the results given by AMF-DFA and AMF-DXA of Stock data sets. To make relation to analysis done in previous subsections, we use Stock market index as input data instead of log-returns data set. Most important results in this regard are as follows:\\

1) The fluctuation function, $\mathcal{F}_{xy}(q;s)$, as a function
of scale, $s$, for different values of $q$ has been computed for
all data sets. All underlying data for time interval used in this
paper behave as power-law with respect to scale, consequently one
can assign scaling exponent, $h_{xy}(q)$. To ensure about the elimination of trends superimposed on data sets, we applied adaptive detrending algorithm. Left panels of  Fig. \ref{figfs1} illustrate original Stock fluctuations for some Stock markets with corresponding trends computed by adaptive detrending method. In this figure, we took ${\rm w_{adaptive}=10}$ ($R={\rm int}\left[\frac{t-1}{11}\right]$), ${\rm w_{adaptive}=50}$ ($R={\rm int}\left[\frac{t-1}{51}\right]$) and ${\rm w_{adaptive}=100}$ ($R={\rm int}\left[\frac{t-1}{101}\right]$) for total number of segments. The higher number of partitioning the better adjustment to original data and consequently, the remaining fluctuation is smoother. The right panels of  Fig. \ref{figfs1} correspond to $\mathcal{F}_{xx}(q=2;s)$ versus $s$. 
Circle symbols in mentioned figure show the results only for regular DFA method while other symbols point out to removing trends using adaptive detrending method. The scaling behaviour of $\mathcal{F}_{xx}(q=2;s)$ for regular DFA is the same as that of given by adaptive detrending for various values of ${\rm w_{adaptive}}$ in small scales. These results confirm that for Stock market fluctuations, regular detrending by DFA and DCCA capable to remove embedded trends  \cite{hu09}. Fig. \ref{figfs}
illustrates $\mathcal{F}_{xx}(q=2;s)$ versus $s$ for some Stock
markets. Since the scaling function for fluctuation function
is justified so we can determine some important exponents to clarify
statistical properties of time series. Generalized Hurst exponent
($h_{xy}(q)$), multifractal scaling exponent ($\tau_{xy}(q)$),
cross-correlation exponent ($\gamma_{xy}$) are some of scaling
exponents used to classify stochastic fluctuation, can be
determined. The upper panel of Fig. \ref{fighq} indicates
$h_{xx}(q)$ as a function of $q$ for some typical series. The
behavior of $\tau_{xx}(q)$ and singularity spectrum,
$f_{xx}(\alpha_{xx})$ USA, Australia, Poland, Italy, Malaysia and India have  been shown in middle and lower panel of Fig.
\ref{fighq}, respectively.
\\
\begin{figure}
\begin{center}
\includegraphics[width=1\linewidth]{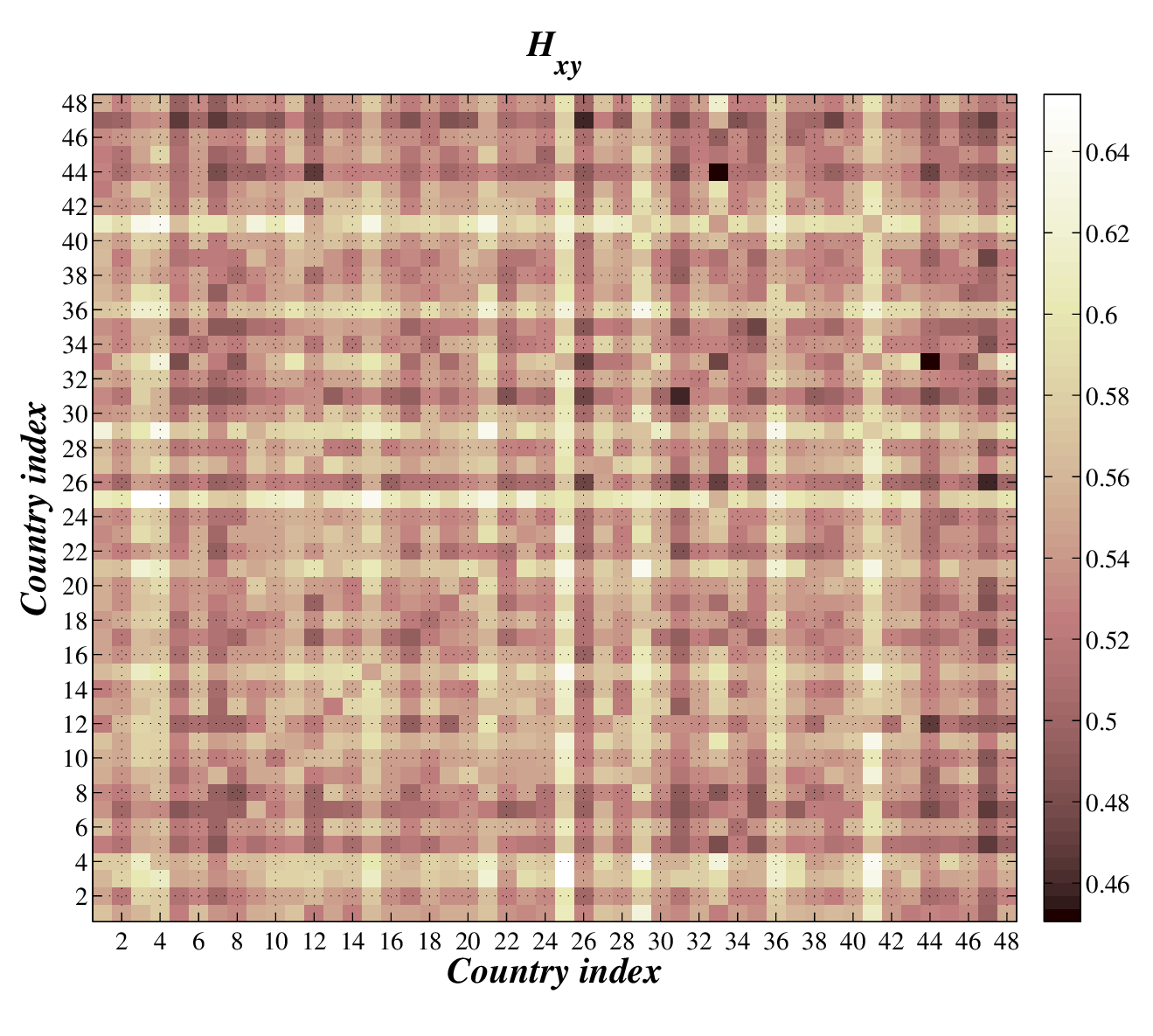}
\includegraphics[width=1\linewidth]{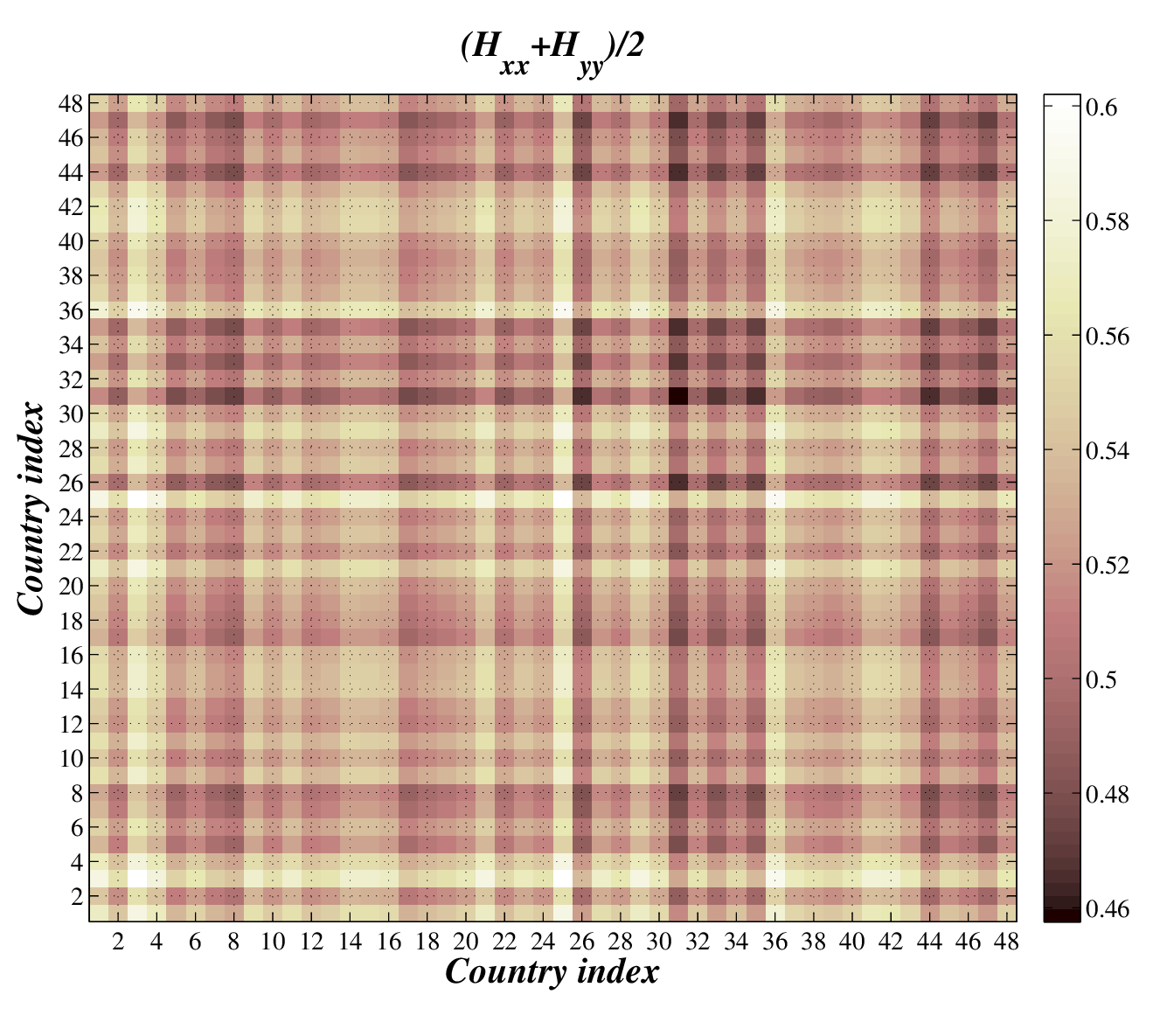}
\caption{\label{fig2} Matrix representation of Hurst exponent ($H_{xy}$) of fluctuation function has been indicted in upper panel. Lower panel shows the average of Hurst exponents of pairs. }
\end{center}
\end{figure}

\begin{figure}
\begin{center}
\includegraphics[width=1\linewidth]{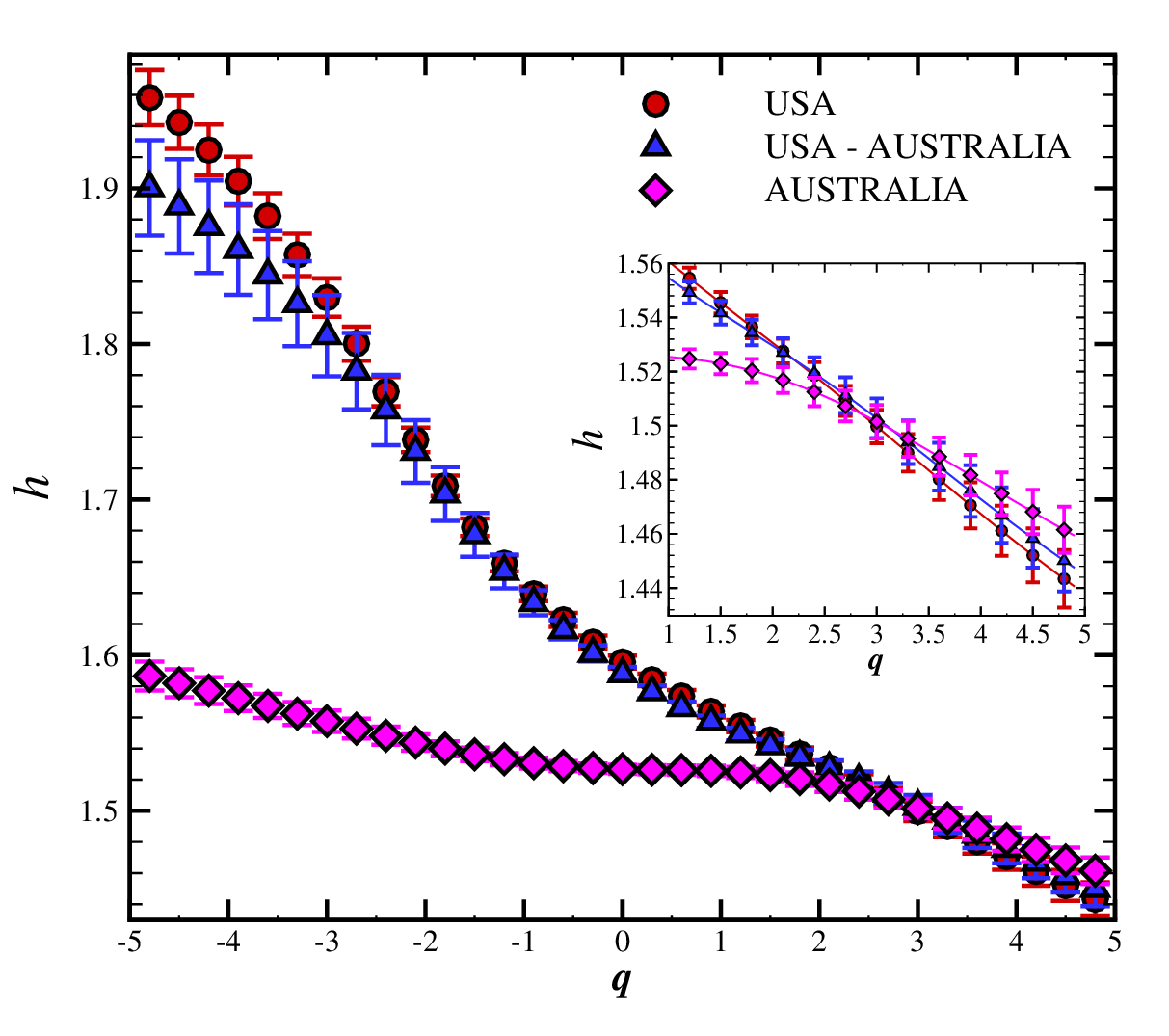}
\caption{\label{fig33} Generalized Hurst exponent of two typical data sets. The relation between generalized hurst exponent determined by MF-DFA and that of given by MF-DXA have been indicated in this plot. For $q\ge 2$ the empirical relation between $h_{xy}$ and $(h_{xx}+h_{yy})/2$ is satisfied while for $q<2$ we have deviation.}
\end{center}
\end{figure}

\begin{figure}
\begin{center}
\includegraphics[width=1\linewidth]{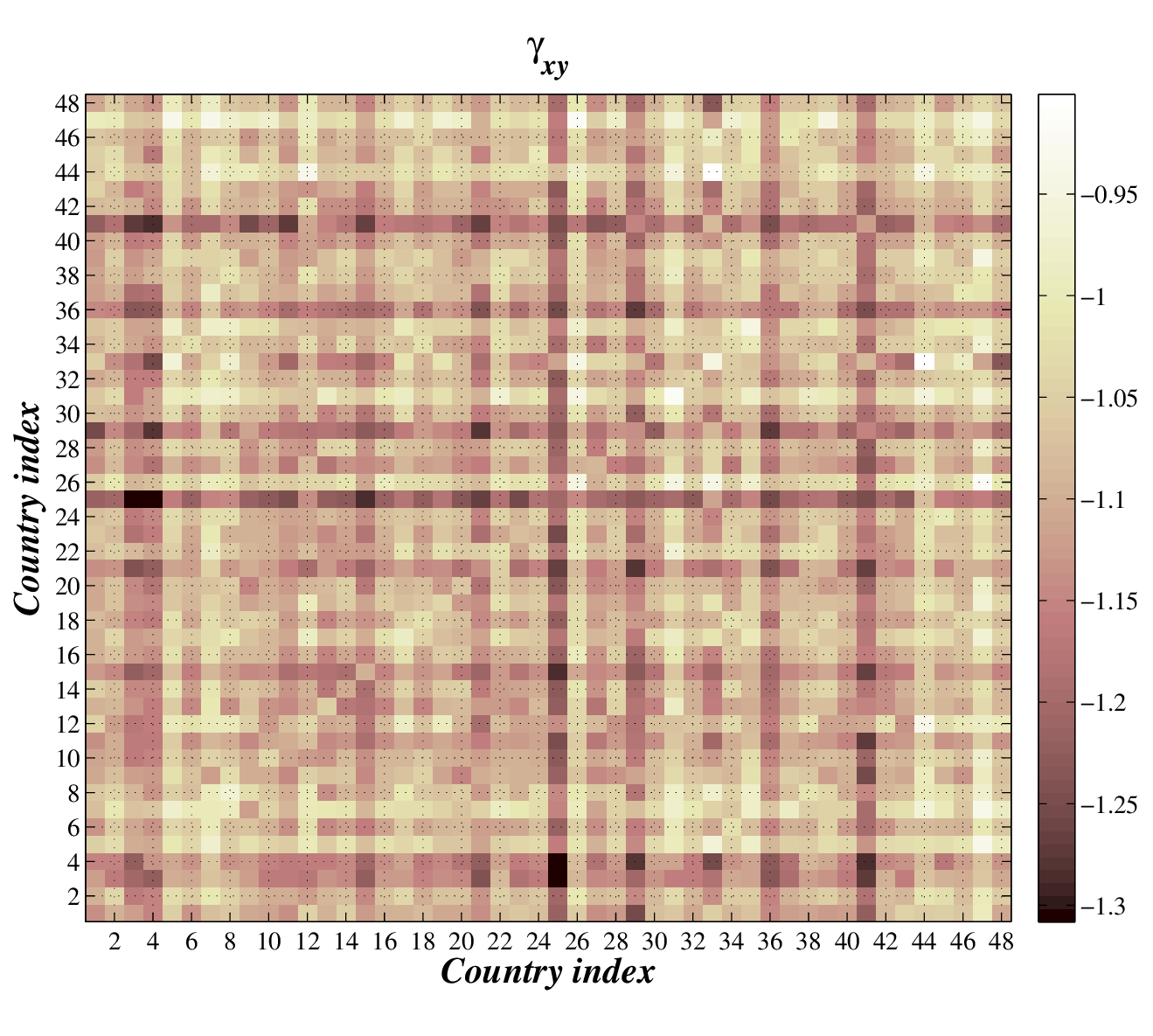}
\includegraphics[width=1\linewidth]{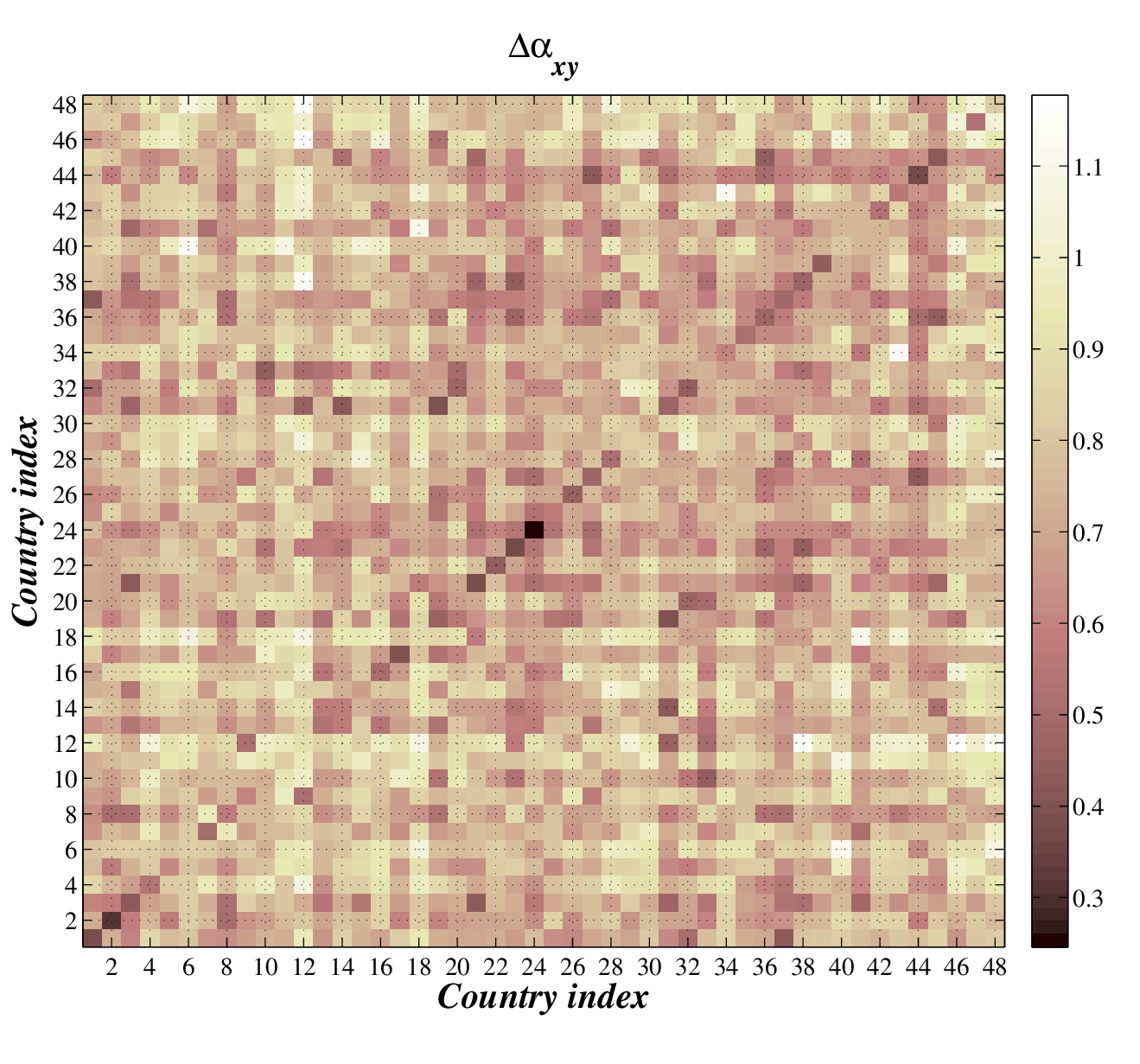}
\caption{\label{fig22} Upper panel corresponds to cross-correlation (off-diagonal elements) and auto-correlation (diagonal elements). $\Delta \alpha_{xy}=\alpha_{xy}^{\rm max}-\alpha_{xy}^{\rm min}$ has been shown in lower panel.}
\end{center}
\end{figure}

2) The value of generalized Hurst exponent for $q=2$ confirms that all underlying data are non-stationary for time interval that we used. So the corresponding Hurst exponent is $H=h(q=2)-1$ (see Table \ref{index}). The range of Hurst exponent is  $H\in [0.457,0.602]$ (see upper panel of Fig. \ref{fig1}).  The index for New Zealand market has lowest Hurst exponent while Jordan index has largest value of Hurst exponent. For $H<0.5$ we have anti-persistent data set. According to  \cite{jafarilevel,vahabilevel} one can conclude that those data sets which have higher value of Hurts exponent belong to emergent markets during the interval used for this study. The value of $\gamma_{xx}$ for all series has been indicated in the middle panel of Fig. \ref{fig1}. \\
3) The $q$-dependency of generalized $h(q)=h_{xx}(q=2)$ for all used series in this research demonstrates multifractality nature of underlying data sets. Fig. (\ref{fighq}) indicates $h_{xx}(q)$ as a function of $q$ for some Stock markets.  To quantify the multifractality nature, we compute $\Delta \alpha_{xy}$ according to Eqs. (\ref{holder1}) and (\ref{holder2}) the strength of multifractality has been reported in Table \ref{index}.  Also lower panel of Fig. \ref{fig1} shows $\Delta \alpha_{xx}$ for 48 data sets. For cross-correlation analysis, the strength of multifractality is $\Delta \alpha_{xy} \in[0.246,1.178]$, consequently, we can conclude that the multifractality in cross-correlation is larger than in auto-correlation (see the lower panel of Fig. \ref{fig22}).  \\

4) The value of $H_{xy}$ as density plot representation in the matrix forms has been shown  in Fig. \ref{fig2}. The diagonal values in these plots correspond to that of indicated in upper panel of Fig. \ref{fig1}.\\

5) Concerning the relation between $h_{xy}(q)$ and $h_{xx}(q)$ and $h_{yy}(q)$, one should state that the empirical relation, $h_{xy}(q=2)\le \frac{h_{xx}(q=2)+h_{yy}(q=2)}{2}$ is satisfied for almost pairs investigated in this research and it is compatible with statement represented in \cite{Ladislav}. For $q>0$, mentioned relation is almost satisfied while there is significant deviation for $q<0$. In other words, for $q<0$ the contribution of small fluctuations in $\mathcal{F}_{xy}(q;s)$ or $\mathcal{F}_{xx}(q;s)$ to be dominated, consequently one can probably conclude that the behavior of markets for small fluctuations is affected by its larger Stock market  pair while for larger fluctuations the cross-correlation contains information from both local (internal) and global (external) conditions and we expect that the empirical relation is satisfied. Fig. \ref{fig33} shows mentioned explanation for two typical countries, namely USA and Australia indices where it seems that small fluctuations in Australia index follows USA index while for larger fluctuations corresponding to more risk phenomena, Australia index takes care also its domestic conditions. \\

6) Fig. \ref{fig22} indicates $\gamma_{xy}$ and the width of H$\ddot{\rm o}$lder namely $\Delta\alpha_{xy}\equiv \alpha_{xy}^{\rm max}-\alpha_{xy}^{\rm min}$. The value of $\gamma_{xy}$ demonstrates that all underlying data sets has mutual interaction. This finding also is confirmed by scaling behavior of $\mathcal{F}_{xy}(q;s)$. Such cross-correlation between series has multifractal nature. Namely, small and large fluctuations have different properties. This finding is relevant for date itself.  As mentioned before, the strength of multifractality is $\Delta \alpha_{xy} \in[0.246,1.178]$, which is larger than that of given in auto-correlation. From statistical physics point of view, this behavior can be linked to increasing the complexity nature of stochastic fields when interactions to be turn on between them. In other words, the broader the multifractality  spectrum, the richer and more complex in structure of fluctuations.  \\

\begin{figure}
\begin{center}
\includegraphics[width=1\linewidth]{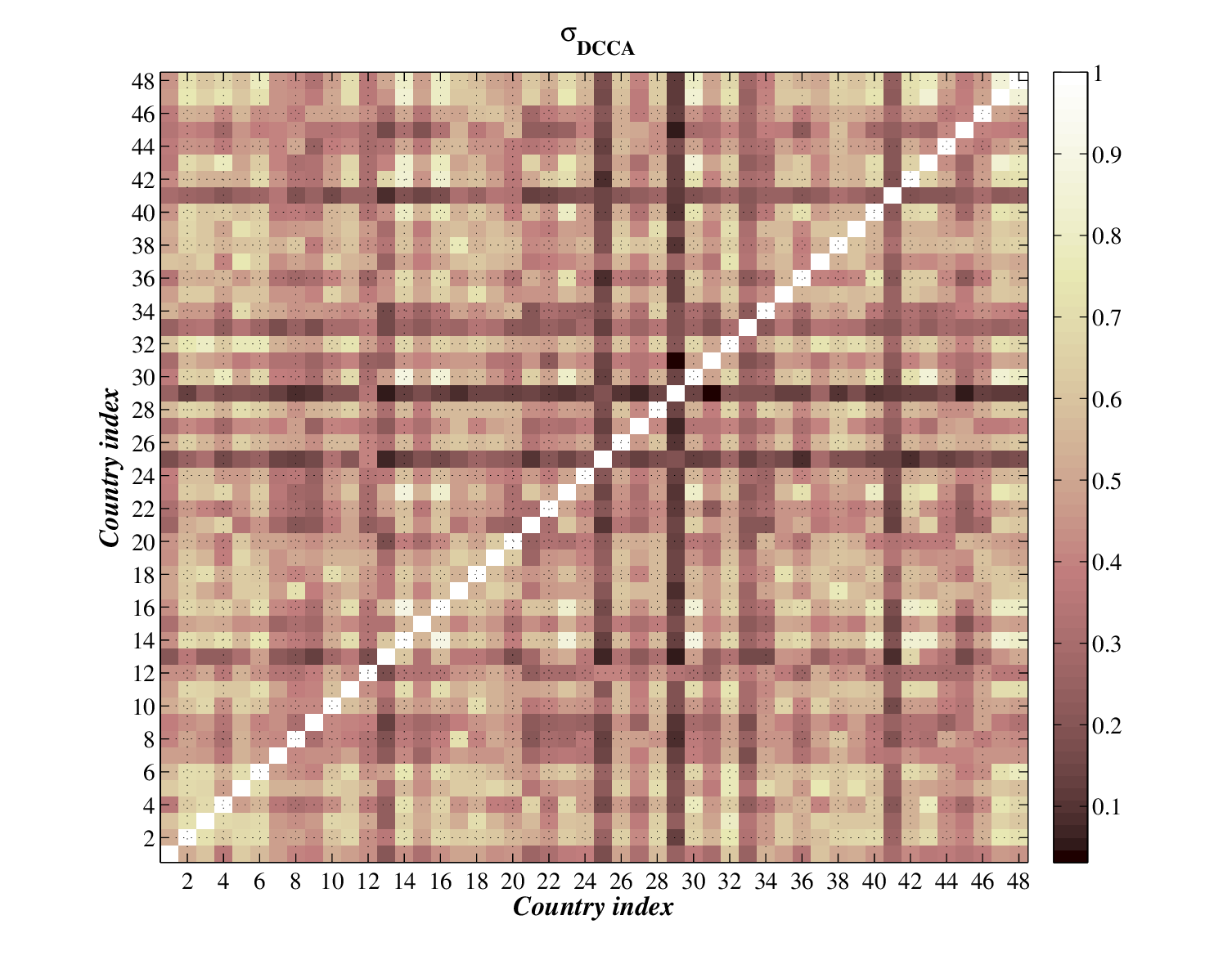}
\includegraphics[width=1\linewidth]{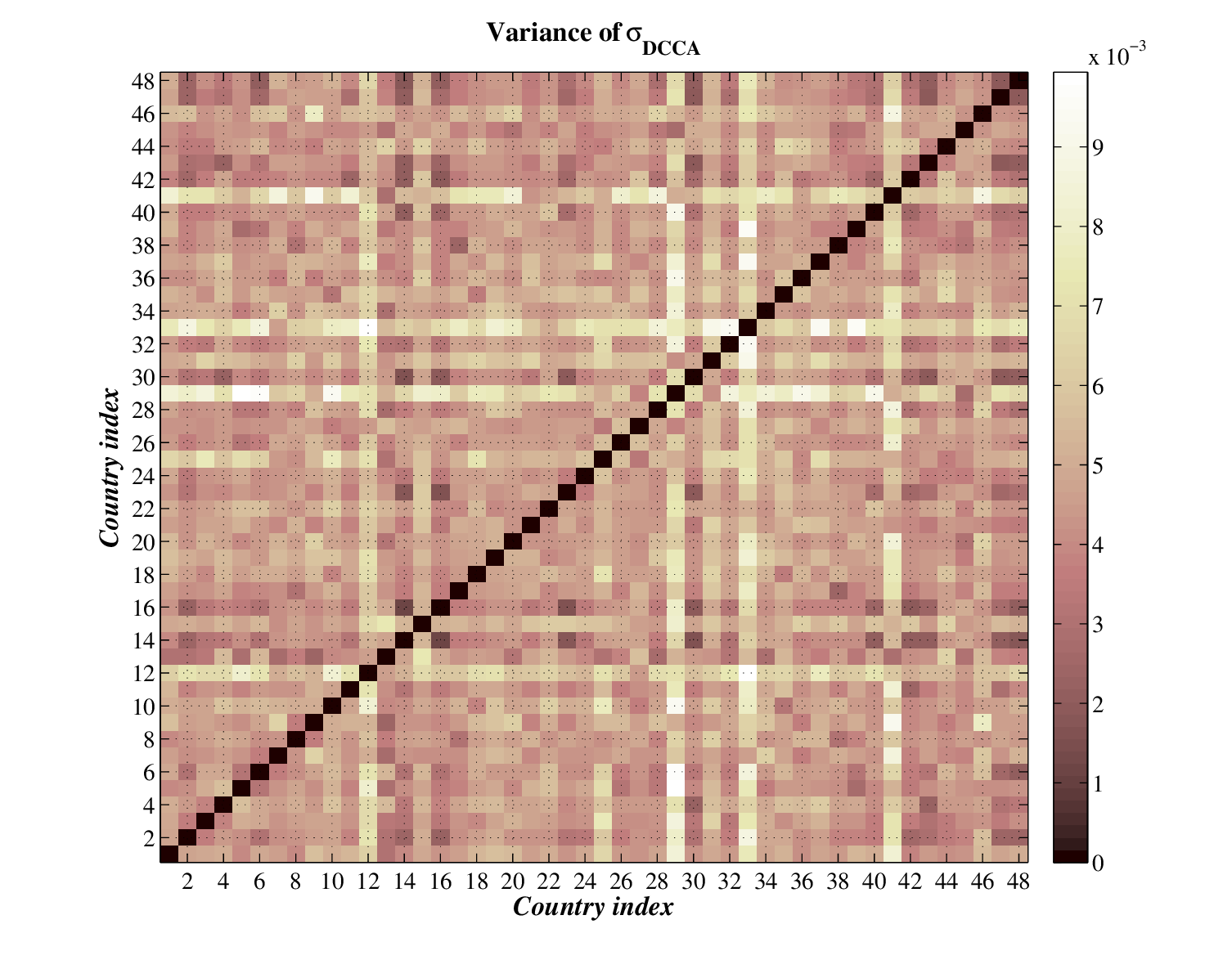}
\caption{\label{fig3} The value of cross-correlation coefficient for countries used in this research has been indicated in upper panel. Lower panel shows the error bars of $\sigma_{xy}$ at $68\%$ confidence interval. The value of $\sigma_{xx}=+1$ so it has no variance, therefore we set the diagonal  elements of error matrix as zero. }
\end{center}
\end{figure}

\begin{figure}
\begin{center}
\includegraphics[width=1\linewidth]{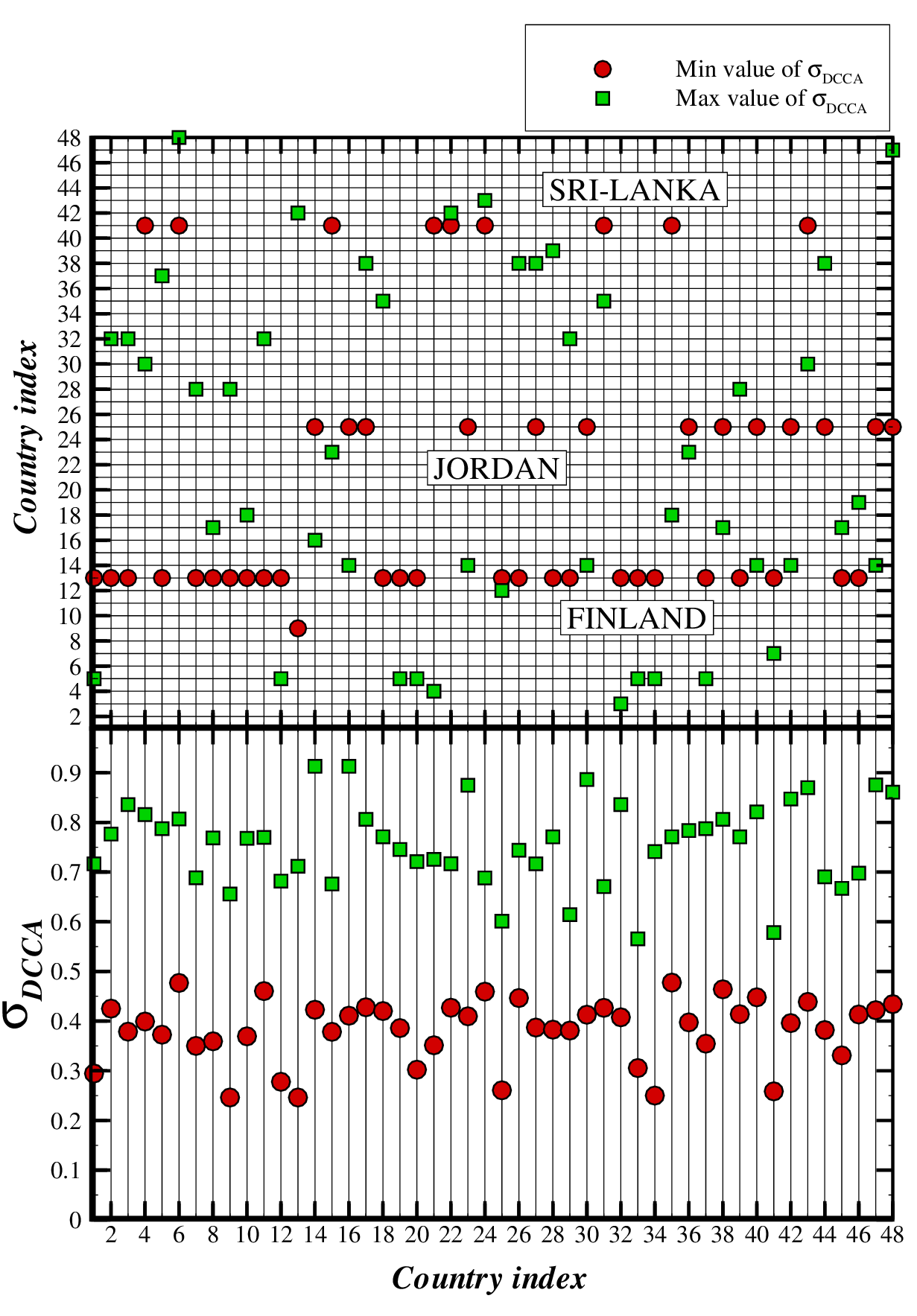}
\caption{\label{minmaxsigma} Upper panel shows the pairs with minimum value of $\sigma_{DCCA}$ (Filled circles) and pairs with maximum value of $\sigma_{DCCA}$ (Filled squares). In lower panel, for each index the minimum and maximum value of $\sigma_{DCCA}$ have been indicated. }
\end{center}
\end{figure}

7) To quantify the nature of cross-correlation between various markets, in addition to compute $\gamma_{xy}$ (upper panel of Fig. \ref{fig22}), we compute $\sigma_{DCCA}$ (see Fig. \ref{fig3}). The interval for this quantity, is $\sigma_{DCCA}\in[0.03,1.00]$. The upper and lower panels of Fig. \ref{fig3} correspond to $\sigma_{DCCA}$ and its variance, $\Delta\sigma_{DCCA}$, respectively. This results demonstrates that all data sets investigated in this paper based on this approach have positive cross-correlation irrespective to the value of their $\gamma_{xx}$'s. The minimum value of cross-correlation based on $\sigma_{xy}$ is for Morocco (Index=29) and New Zealand (Index=31) which is $\sigma_{DCCA}=0.031$. Also, it must point out that Morocco market has very small cross-correlation coefficient with following indices: China, Colombia, Finland, Hong-Kong, Malaysia, New Zealand, Spain and Thailand. Jordan market has small cross-correlation coefficient with Finland, Ireland, Portugal and Sweden indices. Finland market is almost independent from Jordan and Sri-Lanka indices.  The maximum value of cross-correlation coefficient is $\sigma_{DCCA}=0.907$ which is for  France and Germany markets. In Fig. \ref{minmaxsigma} we have plotted the index Stock which has maximum  (filled squares) and minimum (filled circles) values for cross-correlation coefficients In the lower panel of mentioned figure, the minimum and maximum value of $\sigma_{DCCA}$ for each indexes. In these plot we have 3 categories for cross-correlation coefficient based on multifractal approach. First group has minimum cross-correlation with FINLAND. Second group corresponds to  markets have minimum value $\sigma_{DCCA}$ with JORDAN. Third class is devoted to those having minimum $\sigma_{DCCA}$ with SRI-LANKA. According to maximum value of cross-correlation, we can not deduce well-defined classification.\\


\section{Discussions and conclusions}

The application of statistical and mathematical theories are useful to quantify volatilities in economics in order to classify efficient
markets. These tools enable us to collect useful information regarding mutual interactions in Stock markets, risk and optimal portfolio managements in different trading.

Stock market indices are often considered as complex fluctuations
due to the many reasons from economic and social points of view.
Subsequently, applying the common and more trivial methods in data
analysis give incorrect or at least unreliable results. Non-stationarities and unknown noises are also troublesome and disruptive for analysis. Our results
point to some evidence of significant relations between Stock
markets all over the world. Cointegration methods and the Granger
causality show that there are 170 pairs of Stock markets that
exhibit a long-run relationship. These relationships were analyzed
through nonlinear methods, namely mutual information and global
correlation coefficient and the results evidence that are the
emerging and the frontier Stock markets the ones that show the
highest levels of long-run relation and possible predictability. Of
course, we cannot infer immediately that those markets are not
efficient, but we can conclude that the more mature Stock markets
(developed markets) evidence more proximity to the efficiency
hypothesis. These results are complemented with the DFA and DCCA
analysis.

It is worth noting that it has been demonstrated in many previous researches that inferring valuable results for scaling exponents, following necessary conditions should be satisfied: i) The length of recorded series should be large enough and; ii) the probable superimposed trends and noises on the underlying data sets must be small enough or at least distinguishable.  Actually, there is no systematic way to ensure that mentioned conditions to be satisfied, but fortunately, recently some approaches have been developed to pass this bottleneck.

Here,  we rely on one of robust methods in data analysis which is Adaptive Multifractal Detrended Cross-Correlation Analysis (AMF-DXA) 
to explore the mutual effect of Stock market index of 48 countries.
If both input data sets are equal, consequently the results given by
AMF-DXA become the same as that of given by Adaptive Multifractal Detrended
Fluctuation Analysis (AMF-DFA). By applying adaptive detrending algorithm, local trends embedded in data set have been removed. Then clean series have been used for further analysis by MF-DXA. Fig. \ref{figfs1} showed that AMF-DXA can eliminate local trends at reliable level. The value of $h_{xx}(q=2)>1$
demonstrated that all used date in this paper are non-stationary
series. After detrending data set using AMF-DXA method, all relevant
trends embedded in series have been diminished and consequently, a
unique scaling exponent for each $q$'s in all scales was determined
for all pairs (See Figs. \ref{figfs1} and \ref{figfs}). The value of
Hurst exponent which is reported in Table \ref{index} and shown in
the upper panel of Fig. \ref{fig1} represent that, Stock market index for time
interval 2 January of 1995 till 21 January of 2014 belongs to $H\in
[0.457,0.602]$. Therefore some
indices have got anti-persistent and some of them belong to
persistent process and some data sets exhibit uncorrelated nature at
$1\sigma$ confidence interval. According to range of singularity
spectrum, $\Delta \alpha_{xx} =\alpha_{xx}(q_{\bf
min})-\alpha_{xx}(q_{\bf max}) \in [0.304,0.905]$, we concluded that
all underlying data are multifractal (see the lower panel of Fig. \ref{fig1}), while for cross-correlation analysis we found $\Delta \alpha_{xy}=\alpha_{xy}(q_{\bf
min})-\alpha_{xy}(q_{\bf max})\in[0.246,1.178]$ confirming the nature of multifractality in cross-correlation becomes considerable  in comparison with auto-correlation. Subsequently,  cross-correlation causes to increase the complexity nature of behavior of Stock markets.  The empirical relation
between Hurst exponent given by DFA and that of given by DCCA has
been confirmed for data sets. But the relation,
$h_{xy}(q)\le\frac{h_{xx}(q)+h_{yy}(q)}{2}$ is satisfied for almost
pairs investigated in this research just for $q>0$, while there is
significant deviation for $q<0$ (see Fig. \ref{fig33} as an
example).  Since this phenomenon is relevant for almost all pairs
investigated in this paper, we concluded that large fluctuations are
affected by conditions of both underlying markets in each pair, while for
small fluctuations one pair has dominant role and dictates the
behavior of other market in the pair. The value of $\sigma_{DCCA}$ (Eq.
(\ref{dcca1})) determined in this study belongs to the
$\sigma_{DCCA}\in[0.03,1.00]$. The minimum value of this quantity is
for  Morocco (Index=29) and New-Zealand (Index=31) which is
$\sigma_{DCCA}=0.031$. The maximum value of $\sigma_{DCCA}$ is for
France and Germany markets with $\sigma_{DCCA}= 0.907$. This finding confirms the existence of
emergent behavior for underlying markets in time interval used in
this paper.

{\bf Acknowledgements} Paulo Ferreira and Andreia Dion\'{\i}sio are pleased to acknowledge financial support from Funda\c{c}\~{a}o para a Ci\^{e}ncia e a Tecnologia and FEDER/COMPETE (grant PEst-C/EGE/UI4007/2013). S.M.S. Movahed thanks to Alireza Vafaei Sads for his helpful comments in some part of computations. 





\begin{thebibliography}{99}

\bibitem{peter03} Peter F. Christoffersen, {\it Elements of financial risk management}, Academic Press, (2003).  
\bibitem{fama70} E. Fama, The Journal of Finance, Vol. 25, No. 2, pp. 383-417 (1970).
\bibitem{pagan90} A. Pagan, Journal of Empirical Finance, 3(1), 15-102 (1990).
\bibitem{bach900} L. Bachelier, Theory of Speculation, in P. Cootner ed.: The Random Character of Stock Prices (Cambridge, MIT Press, originally published in 1900) (1964).
\bibitem{kenda53}M. Kendall,  Journal of The Royal Statistical Society. 116, 11-25 (1953).
\bibitem{osbo64} M. Osborne, Brownian Motion in the Stock Prices, em P. Cootner ed.: The Random Character of Stock Prices (MIT Press, Cambridge, originally published in 1959) (1964).
\bibitem{morge64} C. Granger and O. Morgenstein, Spectral Analysis of New York Stock Market Prices, in P. Cootner ed.: The Random Character of Stock Prices (MIT Press, Cambridge, originally published in 1963) (1964).
\bibitem{fama63} E. Fama,  Journal of Business. 36 (4), 420-429 (1963).
\bibitem{cont01} R. Cont, Quantitative Finance. I, 223-236 (2001).
\bibitem{camp87} J. Campbell, Journal of Financial Economics. 18, 373-399 (1987).
\bibitem{jafarilevel}G.R. Jafari, S.M.S. Movahed, S.M. Fazeli, M. Reza Rahimi Tabar and S.F. Masoudi, JSTAT, P06008 (2006).
\bibitem{munix12} Michael C. M\"{u}nnix, Takashi Shimada, Rudi Sch\"{a}fer, Francois Leyvraz, Thomas H. Seligman,Thomas Guhr and H. Eugene Stanley, Scientific Reports 2, Article number: 644 (2012).

\bibitem{darb00} G. Darbellay and D.Wuertz, Physica A. 287, 429-439 (2000).




\bibitem{peng92} C.-K. Peng, S.V. Buldyrev, A.L. Goldberger, S. Havlin, F. Sciortino, M. Simons, and H.E. Stanley, Nature, 356, 168-170 (1992).
\bibitem{peng94} C.-K. Peng, S. V. Buldyrev, S. Havlin, M. Simons, H. E.
Stanley, and A. L. Goldberger, Phys. Rev. E 49, 1685-1689 (1994).
\bibitem{bul95} S. V. Buldyrev, A. L. Goldberger, S. Havlin, R. N. Mantegna, M. E. Matsa, C.-K. Peng, M. Simons, and H. E. Stanley, Phys. Rev. E 51, 5084 (1995).

\bibitem{bun02}  
J. W. kantelhardt, S. A. Zschiegner, E. Koscielny-Bunde, A. Bunde, S. Havlin and H. E. Stanley, Physica A {\bf 316}, 87 (2002).
\bibitem{DCCA} B. Podobnik and H. Euge Stanley, Phys. Rev. Lett. {\bf 100}, 084102
(2008).

\bibitem{wu07}Z. Wu et. al., PNAS, 104, 38, p. 14889-14894 (2007).

\bibitem{hu09} Jing Hu, Jianbo Gao and Xingsong Wang, JSTAT, P02066 (2009).

\bibitem{DCCA2} B. Podobnik,  I. Grosse,  D. Horvatic,  S. Ilic,  P. Ch. Ivanov and H. S. Stanley,  Eur. Phys. J. B 71, 243-250 (2009). 

 \bibitem{DCCA3} Boris Podobnik, Zhi-Qiang Jiang, Wei-Xing Zhou, and H. Eugene Stanley, Phys. Rev. E 84, 066118 (2011).


\bibitem{DCCA4} Xi-Yuan Qian, Ya-Min Liu, Zhi-Qiang Jiang, Boris Podobnik, Wei-Xing Zhou, and H. Eugene Stanley, Phys. Rev. E 91, 062816 (2015).
\bibitem{zeb11} G.F. Zebende, Physica A {\bf 390}  614-618 (2011).
\bibitem{Zebende13} G.F. Zebende, M.F. da Silva, A. Machado Filho, PhysicaA, 392 1756-1761 (2013).


\bibitem{pod09} B. Podobnik, Davor Horvatic, Alexander M. Petersen, H. E. Stanley, PNAS December 29,  vol. 106 no. 52, 22079-22084 (2009).



\bibitem{shi14} Wenbin Shi, Pengjian Shang, Jing Wang, Aijing Lin, Physica A 403,  35-44 (2014).
\bibitem{lin09} A. Lin,  P. Shang and X. Zhao,  Nonlinear Dynamics, Volume 67, Issue 1, pp 425-435 (2012).

\bibitem{chen11} L. He and S. Chen,  Chaos, Solitons $\&$ Fractals, 44, pp.355-361 (2011).
\bibitem{ma13} F. Ma, Y. Wei, D. Huang, Physica A, 392, 1659-1670 (2013).
\bibitem{cao14} G. Cao, J. Cao, L. Xu and L. He, Physica A 393, 460-469 (2014).

\bibitem{lin14} A. Lin, P. Shang and H. Zhao, Nonlinear Dynamics, Volume 78, Issue 1, pp 485-494 (2014).
\bibitem{zhao14} X. Zhao, P. Shang and W. Shi, Physica A, 402, 84-92 (2014).

\bibitem{Reboredo14}  J. C. Reboredo,  M. A. Rivera-Castroa,  G. F. Zebende, Energy Economics, Volume 42, March 2014, Pages 132-139 (2014).
\bibitem{Silv15} Marcus Fernandes da Silva, \'{E}der Johnson de Area Le\`{a}o Pereira,
Aloisio Machado da Silva Filho, Arleys Pereira Nunes de Castro, Jos`'{e} Garcia Vivas Miranda, Gilney Figueira Zebende, Physica A, 424 124-129 (2015).
\bibitem {mf-dxa} Wei-Xing Zhou, Phys. Rev. E {\bf 77}, 066211
(2008).


\bibitem{perr92} P. Perron,  T. Vogelsang, Journal of Business and Economic Statistics. 10 (3), 301-320 (1992).
\bibitem{clem98} J. Clemente, A. Monta\~{n}\'{e}s and M. Reyes, Economics Letters, 59. 175-182 (1998).

\bibitem{perr97} P. Perron, Journal of Econometrics. 80 (2), 355-385 (1997).
\bibitem{enge87} R. Engle and W. Granger, Econometrica, 55(2), 251-76  (1987).
\bibitem{joha91}S. Johansen, Econometrica. 59(6), 1551-1580 (1991).
\bibitem{greg96} A. Gregory and B. Hansen, Journal of Econometrics. 70, 99-126  (1996).
\bibitem{greg69} C. W. J. Granger, Econometrica 37 (3): 424-438 (1969).
\bibitem{shan48}C. Shannon, A Mathematical Theory of Communication. Bell Systems Tech. 27, 379-423, 623-656 (1948).





\bibitem{gran04} C. Granger, E. Maasoumi and J. Racine, Journal of Time Series Analysis, 25 (5), 649-669 (2004).

\bibitem{lin94} C. Granger and J. Lin, Journal of Time Series Analysis. 15 (4), 371-384 (1994).

\bibitem{darb98} G. Darbellay,   UTIA Research Report, n. 1889, Acad. Sc., Prague (1998).
\bibitem{soofi97} E. Soofi, Information Theoretic Regression Methods, Fomby, T. and R. Carter Hill ed: Advances in Econometrics - Applying Maximum Entropy to Econometric Problems, 12 (Jai Press Inc., Londres) (1997).
\bibitem{dion06} A. Dion\'{\i}sio,  R. Menezes and D. Mendes, Nonlinear Dynamics. 44, 351-357 (2006).
\bibitem{ferna01} M. Fernandes, Nonparametric Entropy-Based Tests of Independence Between Stochastic Processes. Working Paper (2001).
\bibitem{woo}  Woo Cheol Jun, Gabjin Oh, and Seunghwan Kim, Phys. Rev. E {\bf 73}, 066128 (2006).


\bibitem{kunhu} K. Hu, P. Ch. Ivanov, Z. Chen, P. Carpena and H. E. Stanley,   Phys. Rev. E {\bf 64},
011114 (2001).
\bibitem{trend2} Zhi Chen, Plamen Ch. Ivanov, Kun Hu, H. Eugene Stanley, Phys. Rev. E {\bf 65}, 041107
(2002).

\bibitem{f-dfa}
C. V. Chianca, A. Ticona and T. J. P. Penna,  Physica A {\bf 357},
447 (2005).



\bibitem{f-dfa2} R. Nagarajan and R. G.  Kavasseri,  International Journal of Bifurcation and Chaos, vol.15, no.2, 1767-1773 (2005).

\bibitem{movahedplasma}S. Kimiagar, M. Sadegh Movahed, S. Khorram, S. Sobhanian and M. Reza Rahimi
Tabar, J. Stat. Mech. P03020 (2009).


\bibitem{golub} G. Golub, C. Van Loan, The Johns Hopkins University Press Ltd., London, 1996.

\bibitem{trend3} Radhakrishnan Nagarajan and Rajesh G. Kavasseri, Chaos, Solitons
and Fractals {\bf 26}, 777-784 (2005).

\bibitem{trend3-1} Radhakrishnan Nagarajan and
Rajesh G. Kavasseri, Physica A {\bf 354}, 182-198 (2005).


\bibitem{dccasadegh} S. Hajian, S.M.S. Movahed, Physica A 389, 4942-4957 (2010).



\bibitem{PRL00} A. Bunde, S. Havlin, J. W. Kantelhardt, T. Penzel, J. H. Peter and K. Voigt,  Phys. Rev. Lett. {\bf 85}, 3736 (2000).

\bibitem{taqqu} M. S. Taqqu, V. Teverovsky, and W. Willinger,
 Fractals {\bf 3}, 785 (1995).
\bibitem{sadeghsun} S. M. S. Movahed, G. R.
Jafari, F. Ghasemi, S. Rahvar and M. Rahimi Tabar, J. Stat. Mech,
 P02003 (2006).
 \bibitem{sadeghriver} S. M. S. Movahed and Evalds Hermanis, Physica
A {\bf 387}, 915 (2008).

 \bibitem{zhi11} Zhi-Qiang Jiang and Wei-Xing Zhou, Phys. Rev. E {\bf 84}, 016106 (2011).


\bibitem{muzy94}J.F. Muzy, E. Bacry and A. Arneodo, Int. J. of Bifurcation and Chaos {\bf 4}, 245 (1994).
\bibitem{muzy95}A. Arneodo, E. Bacry and J.F. Muzy, Physica A {\bf 213}, 232 (1994).

\bibitem{halsey86} T. C. Halsey, M. H. Jensen, L. P. Kadanoff, I. Procaccia,
and B. I. Shraiman, Phys. Rev. A {\bf 33}, 1141 (1986).





\bibitem{zhou13} Yu Zhou, Yee Leung and Zu-Guo Yu, Phys. Rev. E {\bf 87}, 012921 (2013).


\bibitem{hosseinabadi12} S. Hosseinabadi, M. A. Rajabpour, M. Sadegh Movahed and S. M. Vaez Allaei, Phys. Rev. E {\bf 85}, 031113 (2012).

 \bibitem{physa} J. W. Kantelhardt, E. Koscielny-Bunde, H. H. A. Rego,
S. Havlin and A. Bunde,  Physica A {\bf 295}, 441 (2001).



\bibitem{cooly65} J. W. Cooley and J.W. Tukey , Mathematics of Computation,
{\bf 19}, 297 (1965).


\bibitem{koscielny98} E. Koscielny-Bunde, H. E. Roman, A. Bunde, S. Havlin and
 H. J. Schellnhuber,  Phil. Mag. B {\bf 77}, 1331 (1998).









\bibitem{dccaq15} J. Kwapie\'{n}, P. Oswiecimka and  S. Drozdz, Phys. Rev. E {\bf 92}, 052815 (2015).







\bibitem{vahabilevel} M. Vahabi, G. R. Jafari, S.M.S. Movahed, J. Stat. Mech.  P11021 (2011).


\bibitem{Ladislav}L. Kristoufek, Physica A, {\bf 431}, 124-127 (2015).

















\end{thebibliography}


\end{document}